\documentclass[journal=jacsat,manuscript=article]{achemso}

\usepackage[version=3]{mhchem} 

\usepackage{graphicx}
\usepackage{dcolumn}
\usepackage{bm}

\usepackage{xcolor}

\author{Harry W. T. Morgan}
\affiliation{Department of Chemistry, University of Manchester, Oxford Road, Manchester M13 9PL, UK}
\affiliation{Department of Chemistry and Biochemistry, University of California, Los Angeles, Los Angeles, CA 90095, USA}
\email{harry.morgan@manchester.ac.uk}

\author{James E. S. Terhune}
\author{Albert Bao}
\author{Ricky Elwell}
\affiliation{Department of Physics and Astronomy, University of California, Los Angeles, CA 90095, USA}

\author{Tyler Kerr}
\affiliation{Department of Chemistry and Biochemistry, University of California, Los Angeles, Los Angeles, CA 90095, USA}

\author{Allana G. Iwanicki}
\author{Tyrel McQueen}
\affiliation{Department of Chemistry, Johns Hopkins University, Baltimore, MD 21218, USA}

\author{Hoang Bao Tran Tan}
\affiliation{Department of Physics, University of Nevada, Reno, Nevada 89557, USA}
\affiliation{Los Alamos National Laboratory, P.O. Box 1663, Los Alamos, New Mexico 87545, USA} 

\author{Udeshika C. Perera}
\author{Andrei Derevianko}
\affiliation{Department of Physics, University of Nevada, Reno, Nevada 89557, USA}

\author{Eric R. Hudson}
\affiliation{Department of Physics and Astronomy, University of California, Los Angeles, CA 90095, USA}
\email{eric.hudson@ucla.edu}

\author{Anastassia N. Alexandrova}
\affiliation{Department of Chemistry and Biochemistry, University of California, Los Angeles, Los Angeles, CA 90095, USA}
\email{alexandrova@g.ucla.edu}

\title[An \textsf{achemso} demo]
  {Investigation into thorium sulfate as a spinless crystal for a high-performance solid-state \ce{^229Th} nuclear clock}

\keywords{}

\begin{document}

\begin{tocentry}
\begin{center}
\includegraphics[width=8cm]{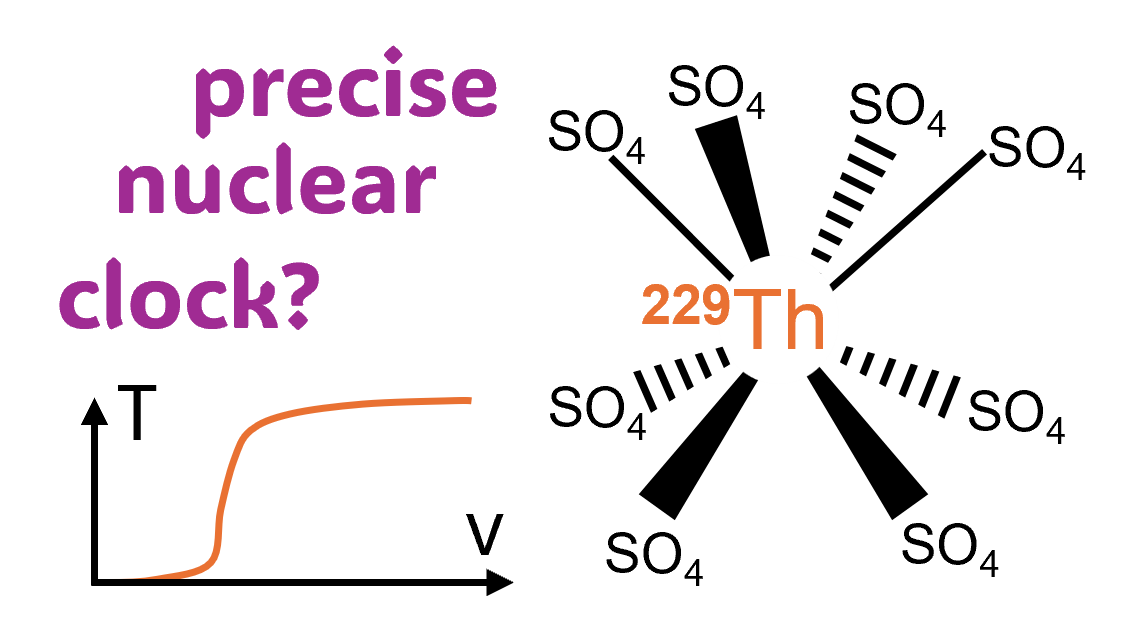}
\end{center}
Thorium sulfate is investigated as a possible \ce{^229Th} solid-state nuclear clock material through synthesis, spectroscopy, and quantum chemistry.

\end{tocentry}

\begin{abstract}
The \ce{^229Th} isotope has a laser-accessible nuclear transition allowing the construction of a solid-state nuclear clock.
Solid-state \ce{^229Th} nuclear clock development has focused on \ce{^229Th}-doped \ce{CaF2} and other metal fluorides as the solid material.
However, the accuracy of any fluoride-based clock will be reduced by nuclear magnetic dipole coupling to \ce{^19F} and by inhomogeneities in the thorium defect environments.
Here we explore thorium sulfate, \ce{Th(SO4)2}, as a solution to both problems.
Strong bonding and charge delocalization in the \ce{SO4^2-} anion may give \ce{Th(SO4)2} the wide band gap needed for a good clock.
In this work we synthesize \ce{Th(SO4)2} and probe it as a nuclear clock material spectroscopically, finding that it is opaque to the 148 nm radiation that excites the \ce{^229Th} nucleus.
Theoretical analysis of the optical spectra shows that charge transfer excitons are responsible for the absorption.
We give guidelines for future material development and assess the prospects for \ce{Th(SO4)2} as a clock material.
\end{abstract}


\section{Introduction}
The \ce{^{229}Th} isotope has a nuclear transition that can be driven with a laser when the \ce{^{229}Th} is in a solid-state material~\cite{RN578}.
It is analogous to a M\"ossbauer transition but with a much lower energy, longer lifetime, and narrower linewidth.
It should allow the construction of a robust and portable solid-state nuclear clock~\cite{RN604} with performance far beyond caesium atomic clocks.
The impacts of a solid-state nuclear clock are expected to span navigation and communication technology, fundamental physics\cite{RN632, RN578}, and even prediction of earthquakes.\cite{RN573,RN574}
Excitement about this has erupted since the demonstrations of excitation of the \ce{^{229}Th} nucleus with a 148 nm (8.4 eV) laser in \ce{CaF2}, \ce{LiSrAlF6}, and \ce{ThF4}~\cite{RN544, RN602, RN629,Derevianko2026Colloquium}, and demonstrations of feedback-locked clock systems with \ce{CaF2}.\cite{toscanidecol2026clock,Huang2026clock}

The clock works by tuning a laser to be resonant the \ce{^229Th} nuclear transition and then counting individual waves of light with an electronic detector; a certain number of counts define a fixed period of time.
The SI unit system defines the second in this way, with a hyperfine transition in a \ce{^133Cs} atom as the light source (\textit{i.e.} an atomic clock).
Tuning the laser requires detection of the excited state, denoted \ce{^{229m}Th}, which can be done in several ways.
Most clock development efforts to date have detected \ce{^{229m}Th} optically, through absorption or fluorescence spectroscopy.
Optical detection has the advantage of minimizing the homogeneous (lifetime) broadening, but measurements are slow because of the long excited state lifetime and the precision may be limited by inhomogeneous (environmental) broadening.
An alternative is to detect electrons produced when \ce{^{229m}Th} decays non-radiatively,\cite{Elwell2025ThO2} which has a larger homogeneous linewidth but faster measurements that allow for quicker data collection, giving better performance in real-time applications like high-speed navigation, and possibly lower inhomogeneous broadening than optical materials.


In optical detection systems, the crystal containing \ce{^229Th} must have a band gap larger than the nuclear transition energy of 8.4~eV when doped with \ce{^229Th} to allow excitation of the nucleus.
It is also desirable that the introduction of \ce{^229Th} into the material does not create a low-lying electronic state that might quench the nuclear excited state non-radiatively~\cite{our_photoquenching, PTB_photoquenching, IC_rate_theory, Pineda2025}.
The current paradigm for achieving this is $s$-block metal fluorides ~\cite{RN459,RN460,RN575,RN544,RN531,RN532,RN552,RN543,RN602,RN629,BaMgF4}, because fluorine is the most electronegative element~\cite{RN344}.
However, the inhomogeneous broadening from dipolar coupling between the \ce{^229Th} nucleus and the surrounding \ce{^19F} nuclei reduces clock accuracy by up to six orders of magnitude.\cite{RN578,RN629}
While that could possibly be mitigated by techniques like magic-angle spinning~\cite{RN607} and dynamical decoupling~\cite{dynamical_decoupling}, it would be advantageous to remove it entirely.
Furthermore, introducing thorium as a dopant causes problems with inhomogeneity of the thorium environment and microscopic strains in the crystal, which cause further inhomogeneous broadening.\cite{RN578}
It also makes characterization of the thorium environment very difficult.\cite{Hiraki2025LaserMossbauer}
There is therefore a strong motivation to develop stoichiometric thorium compounds as nuclear clock hosts.

Here we replace fluoride with a spinless anion, namely the sulfate anion, \ce{SO4^{2-}}. 
S and O both have predominant zero-spin isotopes, but thorium dioxide and disulfide have small band gaps~\cite{Guo2016ThS2,Griffiths2000ThO2,Pereira2017ThO2}.
We propose \ce{SO4^{2-}} as a solution to this problem~\cite{RN571,RN572,RN577,RN601,Morgan2025polyatomic}.
It has strong S-O bonds and exhibits delocalization of the negative charges, so electron transfer from \ce{SO4^{2-}} to a metal cation like \ce{Th^{4+}} has a large energy cost.
\ce{Th(SO4)2} could therefore combine the favourable nuclear properties of \ce{O} and \ce{S} with the electronic properties needed for a large band gap and, as \ce{Th(SO4)2} is a stoichiometric material, there is no interference from charge-compensating defects~\cite{RN575,RN460,RN531,RN611,BaMgF4}.

The electron detection clock system has been demonstrated using \ce{ThO2},\cite{Elwell2025ThO2} which avoids the problems of nuclear spin and doping but suffers accuracy losses because the \ce{^{229m}Th} non-radiative decay is too fast.
To improve that we need to tune the decay rate \textit{via} the electronic structure.\cite{Perera2026AbInitio}
\ce{Th(SO4)2} could also achieve that without introducing new sources of error.

In this work we assess \ce{Th(SO4)2} as a nuclear clock material experimentally and interpret the results with theoretical support.
We synthesize crystalline \ce{Th(SO4)2} and perform VUV transmission spectroscopy to see whether the compound is transparent to 148 nm radiation.
We then perform state-of-the-art electronic structure calculations to predict the optical spectra and analyse the electronic structure of thorium.

\begin{figure}
    \centering
    \includegraphics[width=0.5\linewidth]{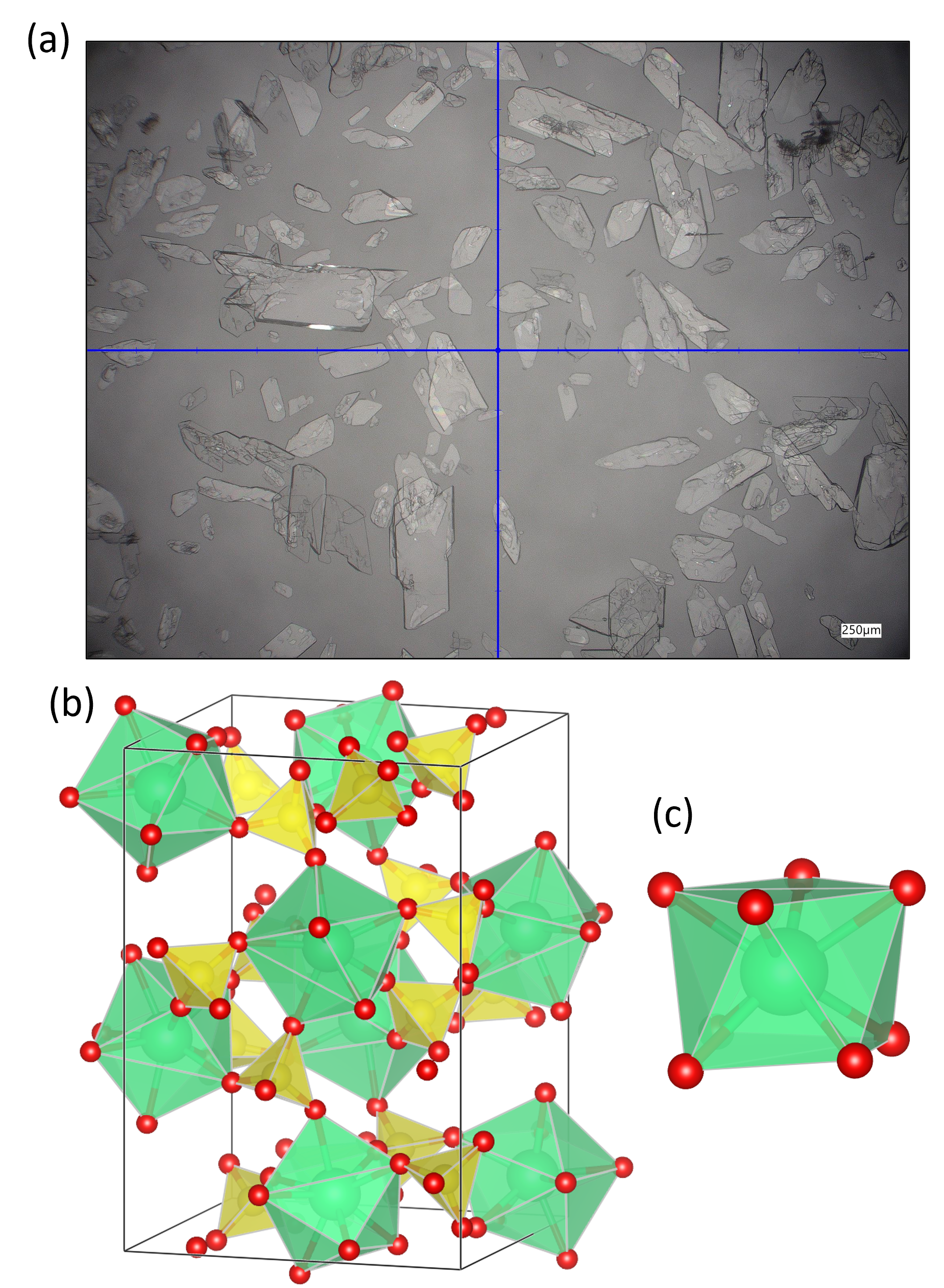}
    \caption{(a) microscope image of \ce{Th(SO4)2} crystals. (b) Unit cell of \ce{Th(SO4)2}. All Th atoms are related by symmetry. (c) \ce{ThO8} coordination shell. Th atoms are shown in green, S in yellow, and O in red.}
    \label{fig:ThSO42 structure}
\end{figure}

\section{Results}
\subsection{Synthesis}
While preliminary synthesis and characterization of  \ce{Th(SO4)2} had been reported in the literature\cite{Betke2011}, the crystals had not yet been grown in sufficient purity and size for optical characterization. In the literature synthesis, \ce{Th(SO4)2} crystallized alongside additional phases containing co-crystallized hydrogen sulfate solvent molecules. We optimized this growth, using principles of high temperature chemistry, by considering entropy and accordingly increasing the temperature relative to the literature synthesis.
As temperature is increased, it is expected that the system will gain entropy by releasing solvent molecules from coordination complexes back into solution. We predicted and experimentally found that we were able to select only the ansolvate \ce{Th(SO4)2} phase by performing the synthesis at 350$^{o}$C. This growth strategy is likely to be fruitful in the synthesis of other novel ansolvate phases, which is important for making nuclear-spin-free materials.

\ce{Th(SO4)2} was synthesized solvothermally in a evacuated quartz tube. 150 mg of Th(NO$_{3}$)$_{4}$*4H$_{2}$O (United Nuclear) powder was added to a 30 cm 6x8 fused quartz tube and combined with 2 mL of concentrated \ce{H2SO4} (JT Baker, 96.7\%, Lot \#K46062), using the solvent to wash the powder down the sides of the tube. The tube was cooled to 77 K in a dewar flask of liquid \ce{N2} for 5 minutes and necked 20 cm from the bottom without any melting of the solution. The tube was re-cooled for five minutes before placing it on a vacuum line, where it was flushed with argon three times before sealing under vacuum. The tube was immediately placed back in the dewar flask to cool, then once frozen was removed and allowed to reach room temperature on the benchtop. 

The sample was heated upright at an angle in a box furnace at 350$^{o}$C for 3 days with 100$^{o}$C/hr heating and cooling rates. Slow-cooling, a longer dwell time, a less concentrated solution, horizonal growth in a temperature gradient, and iterative growths using seed crystals all yielded crystals of the same size and morphology as the original growth. This suggests crystal growth happens at 350$^{o}$C and occurs concurrently with compound formation. 

After synthesis, the tube was opened in the fume hood and transferred to a solvent for transmission measurements. The excess solvent was decanted and the crystals were washed by adding and decanting sulfuric acid three times, then sulfolane three times, then ethanol.  The crystals were found to be stable for approximately 1 year in the growth medium and in sulfuric acid, but decomposed slowly in other solvents. Clear plate-like crystals ranging from 0.1 to 1 mm were obtained by this method.

\subsection{Spectroscopy}
The \ce{Th(SO4)2} crystal used for spectroscopy, approximately 600 $\times$ 250 $\times$ 10 microns in size, was first washed with isopropyl alcohol to eliminate any residual VUV absorbing solvents. It was then loaded into a custom mount and covered with a \ce{CaF2} coverslip, where it was then placed inside a Seya-Namioka vacuum monochromator (McPherson 302). Broadband VUV light, produced by a deuterium lamp (McPherson 632), travelled through our crystal and onto a diffraction grating. The grating was typically run from 140 nm to 250 nm in 0.5 nm increments. Photon counts were detected by a CsI photo-multiplier tube (Hamamatsu R6836) and counted by a multi-channel scalar (Stanford Research Systems SR430). Additional details on sample preparation can be found in the supplementary information.

\begin{figure}[h]
    \centering
    \includegraphics[width=0.5\linewidth]{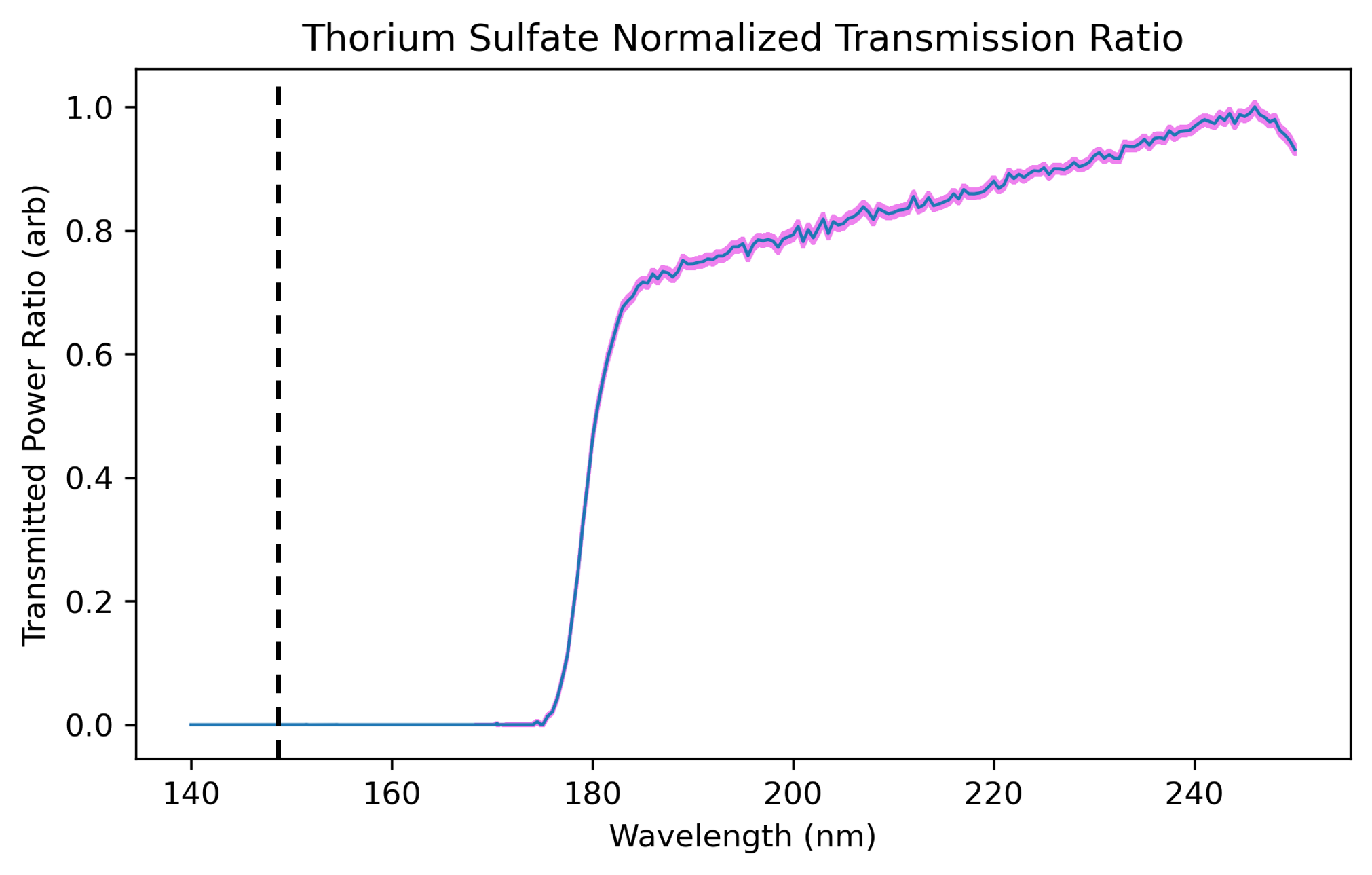}
    \caption{Experimental transmission spectrum of \ce{Th(SO4)2}. 90\% CI errors shown in violet. The vertical dashed line represents the nuclear transition energy (148 nm).}
    \label{fig:Sulfate_transmission}
\end{figure}

The transmission was calculated as the normalized ratio of photon counts with and without the sulfate crystal loaded.
Figure \ref{fig:Sulfate_transmission} shows the resultant transmission ratio, with the 90\% confidence interval Poissonian errors shown in violet.
Efficient transmission was only seen at wavelengths longer than 180 nm; between 180 nm and 175 nm, the transmission dropped to approximately zero. Below 175 nm, the transmission is zero. Transmission measurements were repeated on three different crystals of similar dimensions, all of which gave consistent results. Single-crystal X-ray diffraction was performed on multiple sample crystals, all of which verified the identity of our samples as \ce{Th(SO4)2}. Diffraction data can be found in the supplementary information.


\subsection{Electronic structure calculations}

The most accurate predictions of optical properties come from the \textit{GW} approach, an approximate many-body theory more suitable for excited state properties~\cite{RN624,RN625,RN595,RN489,RN628}.
For accurate optical property predictions for \ce{Th(SO4)2} we use $G_0W_0R$, a variation of the more common $G_0W_0$ method with reduced system size scaling and higher memory requirements~\cite{RN626,RN627}.
This calculation yields a direct band gap of 8.18 eV (152 nm) at the $\Gamma$ point, just below the 8.4 eV nuclear transition energy.
An earlier report from this investigation contained a predicted band gap of 9.0 eV; this was due to an error in the processing of the $G_0W_0$ data
.\cite{Morgan2025ThSO42}

\begin{figure}[htb]
    \centering
    \includegraphics[width=0.5\linewidth]{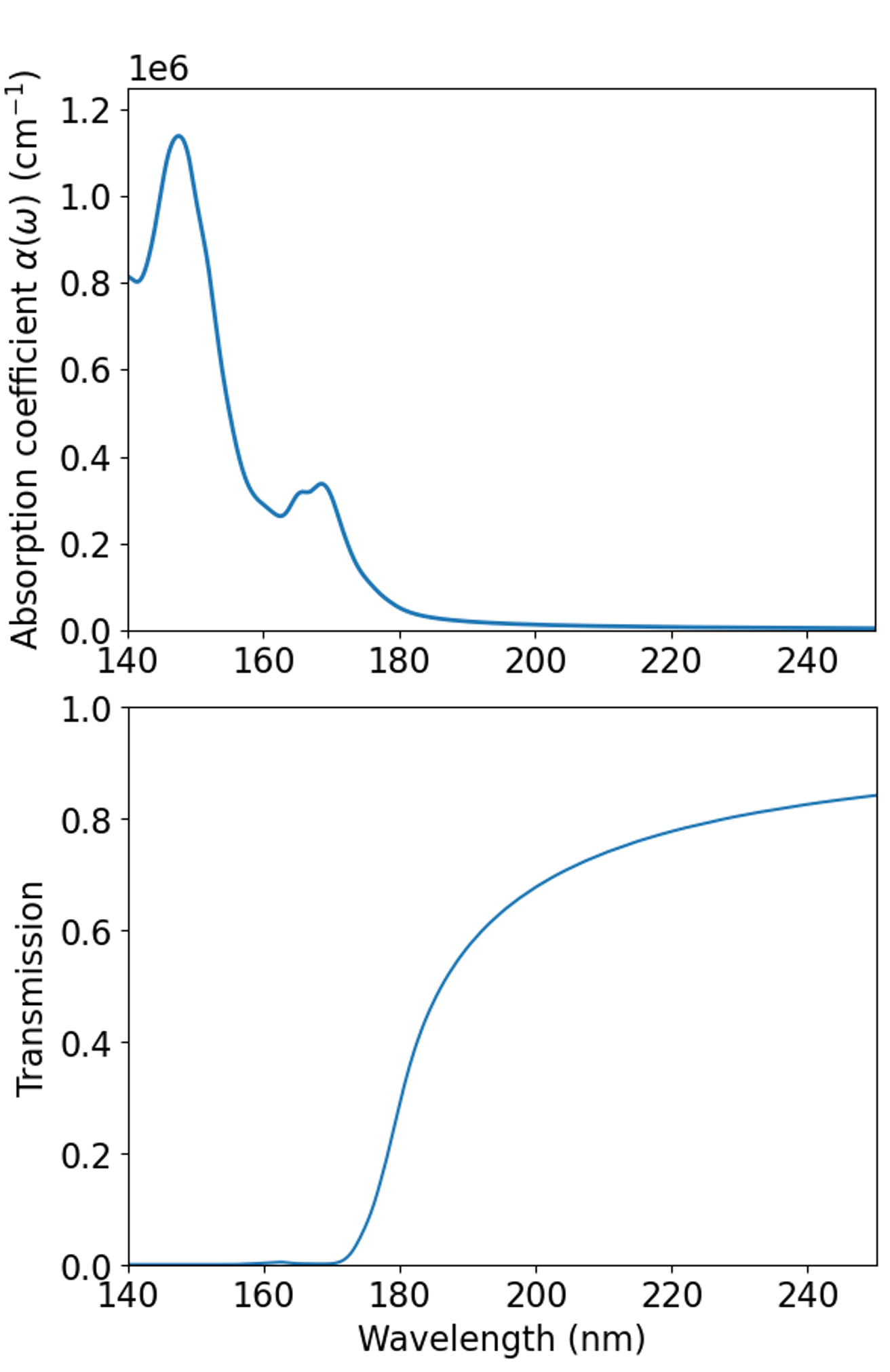}
    \caption{Optical absorption and transmission spectra of \ce{Th(SO4)2} computed with $G_0W_0R+$BSE.}
    \label{fig:GW+BSE absorption spectrum}
\end{figure}

We then used the Bethe-Salpeter Equation (BSE) method to compute the dielectric function and optical spectra including excitonic effects (bound electron-hole pairs).
The calculated absorption and transmission spectra are shown in Figure \ref{fig:GW+BSE absorption spectrum}.
The transmission spectrum was computed by:
\begin{align}
    T = \frac{(1-R)^2e^{-\alpha d}}{1-R^2e^{-2\alpha d}}
\end{align}
where $R$ is the reflectivity, also computed with $G_0W_0+$BSE, and $d$ is the sample thickness.\cite{Fox2010optical}
This formula accounts for internal reflections, though that is a negligible effect here.
A value of $d= 0.2\;\mu$m was used, a factor of 50 smaller than the experimental sample thickness.
The match between the computed and experimental transmission spectra are excellent, but why a smaller $d$ parameter is needed is not clear.
Further investigation can be found in the SI.

The first absorption peak appears at $\sim 170$ nm, matching the experimentally observed absorption.
It is significantly below the band gap, so it corresponds to excitonic transitions.
We conclude that the excitonic transitions completely absorb the incident light in this system.

\begin{figure}[tb]
    \centering
    \includegraphics[width=0.5\linewidth]{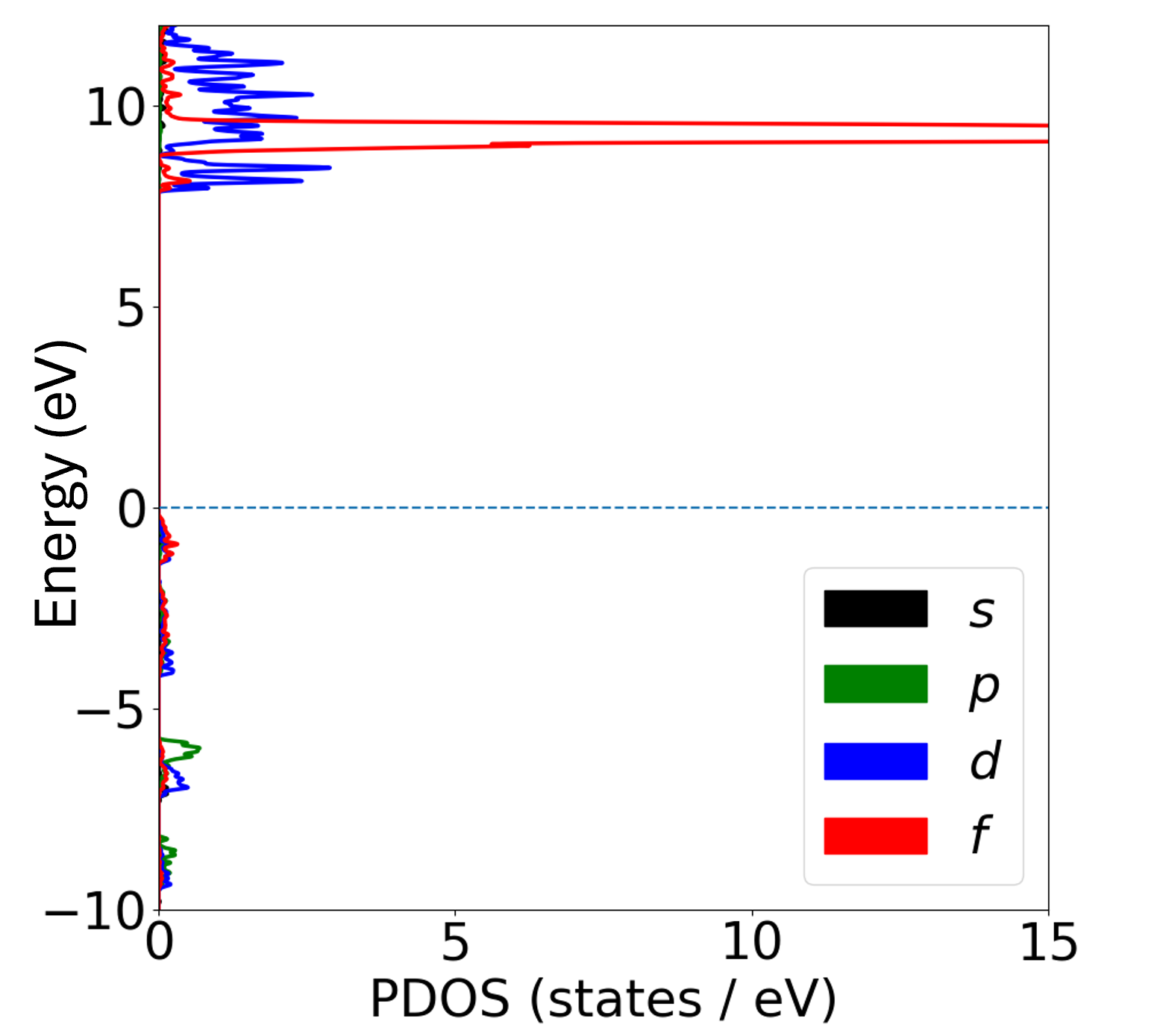}
    \caption{Orbital-projected density of states (PDOS) for thorium in \ce{Th(SO4)2} calculated with $G_0W_0R$}
    \label{fig:Th PDOS}
\end{figure}

To understand the electronic structure of \ce{Th(SO4)2} we plot the thorium projected density of states (PDOS) as calculated by $G_0W_0R$ (Figure \ref{fig:Th PDOS}).
The valence band contains very small contributions from thorium, as expected for closed-shell \ce{Th^4+}.
The conduction band consists of a broad $6d$ band and a sharp $5f$ peak.
This is different from results for other candidate \ce{^229Th} clock materials in that the conduction band minimum is Th $6d$, not $5f$.\cite{Morgan2025polyatomic,Nickerson2021,RN575}
This is probably a result of superior treatment of $5f$ electron correlation by $GW$ and $6d$ orbital overlap effects (\textit{vide infra}).

To characterize the excitons we plot the electron densities associated with the strongest transitions that contribute to the absorption peak at $\sim7$ eV.
The hole is localized on an O atom.
Further details of this visualization are given in the supplementary information.
The most intense transitions are computed at 170.3 nm, 170.1 nm, 168.9 nm, and 168.5 nm.

\begin{figure}[htb]
    \centering
    \includegraphics[width=0.45\linewidth]{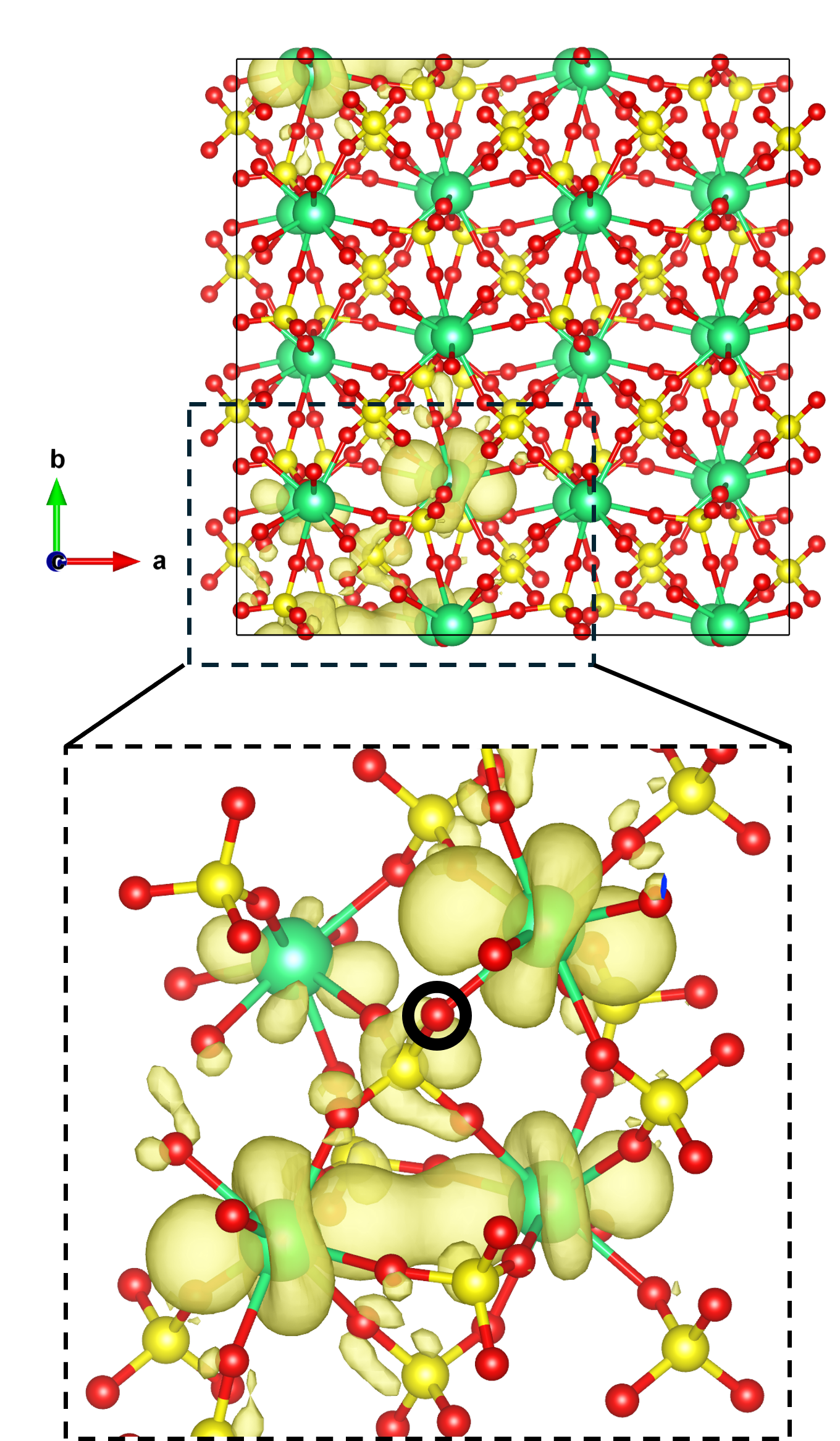}
    \caption{Excitonic electron density for the 168.5 eV transition in \ce{Th(SO4)2} with the hole fixed near an O atom. Th atoms are represented by green spheres, S by yellow spheres, O by red spheres, and the electron density by the yellow surface. The black circle in the insert marks the position of the hole.}
    \label{fig:exciton 56 charge density}
\end{figure}

Figure \ref{fig:exciton 56 charge density} shows the excited electron density in the 168.5 nm transition.
It has Th $d$ orbital character spread over 4 Th atoms, so it could be described as a multicentre ligand-to-metal charge transfer (LMCT) transition, though some electron density on a central sulfate ion is also visible.
Corresponding plots for the other three transitions are given in the SI, all showing Th $d$ character on the same group of 3-4 Th atoms.
The absorption onset can therefore be ascribed to multicentre LMCT-like transitions.
LMCT transitions in transition metal complexes are typically strong absorbers because they are allowed by selection rules, which is consistent with our experimental observations here.
Absorption peaks at 180-195 nm are observed in aqueous solutions of sulfates and sulfuric acid.\cite{Cheng2022,Cheng2023,Ai2021,Ai2022}
However, the environment of \ce{SO4^2-} in a dilute aqueous solution is very different from that in \ce{Th(SO4)2}, and in concentrated solutions the absorption has been attributed to charge transfer from \ce{H2O} to \ce{H2SO4},\cite{Cheng2023} so it is reasonable to attribute the absorption in \ce{Th(SO4)2} to a different mechanism involving the \ce{Th^4+} cations.

The electron densities for the 170.3 nm, 168.9 nm and 168.5 nm transitions (the latter in Figure \ref{fig:exciton 56 charge density}) also show substantial overlap between the $d$ orbitals of neighbouring thorium atoms.
This overlap explains the $d$ peaks below the main $f$ band in the Th PDOS (Figure \ref{fig:Th PDOS}).
Similar interactions were seen in $G_0W_0$ calculations on \ce{ThF4} and \ce{Na2ThF6}.\cite{RN543}
However, the Th-Th pairs showing $d$ overlap in \ce{Th(SO4)2} have a Th-Th distance of 5.03 \AA{}, too long for a covalent-bonding-based description of the excitons according to covalent radii.\cite{Pyykk2014}
\ce{SO4^2-} may therefore have a role in stabilizing the exciton.

The characterization of the excitons shows the way to expand the optical transparency window.
The excitons responsible for the absorption onset are stabilized by Th-Th orbital interactions, so they ought to be destabilized by increasing the Th-Th distances.
That could be done by co-crystallizing the thorium sulfate with solvent, as in hydrated crystals, or other anions, as in \ce{Th(HSO4)2(SO4)}.\cite{Betke2011,RN597,Knope2011}
Removing nuclear spins completely from such systems will be a challenge, but the reduction of nuclear spin relative to fluorides may offer better clock stability.


Finally, there is an alternative clock system to the optical detection of \ce{^{229m}Th} that could work with \ce{Th(SO4)2}, namely conversion electron detection.
Since the \ce{Th(SO4)2} band gap lies just below the nuclear transition energy, the ``internal conversion'' (IC) relaxation process is energetically allowed.
IC in a crystal occurs when \ce{^{229m}Th} relaxes by promoting a valence band electron to the conduction band rather than emitting a photon.\cite{Morgan2025ICrate,Perera2026AbInitio}
Note that this is unrelated to the ``internal conversion'' of molecular spectroscopy.

The key parameter for clock performance in this mode is the IC lifetime, $\tau_{\rm IC}$, because a longer lifetime gives a lower homogeneous linewidth and a more precise clock.
We can compare \ce{Th(SO4)2} to \ce{ThO2}, in which electronic detection of \ce{^{229m}Th} has already been done.\cite{Elwell2025ThO2}
We can estimate the IC lifetime using the formalism of \citet{Perera2026AbInitio} by rigidly shifting the band gap of \ce{ThO2} up to the value for \ce{Th(SO4)2} computed here, since both compounds contain \ce{Th^4+} ions.
That gives an IC rate for \ce{Th(SO4)2} of $162~\mu\mathrm{s}$, much longer than $\sim10 \;\mu$s in \ce{ThO2}, reducing the associated linewidth, $\Delta\nu_{\rm IC}\simeq(2\pi\tau_{\rm IC})^{-1}$, from approximately $16~\mathrm{kHz}$ to $1~\mathrm{kHz}$.
\ce{Th(SO4)2} is therefore an attractive candidate for a solid-state nuclear clock with an electron current readout.
Accurate prediction of $\tau_{IC}$ for \ce{Th(SO4)2} will require inclusion of excitons in the IC theory.

\section{Conclusions}

We have assessed \ce{Th(SO4)2} as a nuclear clock material with synthetic, spectroscopic, and theoretical methods.
\ce{Th(SO4)2} contains no spin-active nuclei, allowing clock stability to be increased by orders of magnitude over existing fluoride systems.
Refinement of an established synthetic process allowed us to produce crystals of \ce{Th(SO4)2} suitable for optical spectroscopy with control over the amount of co-crystallized solvent.
Comprehensive spectroscopy was performed on three crystal samples, with careful consideration given to eliminating the influence of extraneous VUV absorption sources, ensuring that the measured spectra reflects the intrinsic \ce{Th(SO4)2} transmission.
These measurements found that \ce{Th(SO4)2} strongly absorbs wavelengths shorter than 180 nm so it is not useable as a nuclear clock material in an optical readout mode.
$GW+$BSE calculations show that this absorption is due to excitons in which \ce{SO4^2-} lone pair electrons are transferred to \ce{Th^4+} $d$ orbitals, and that the excitons are delocalized over 3-4 \ce{Th^4+} centres.
The optical transparency window may therefore be widened by designing materials with larger Th-Th separations to disrupt this stabilizing delocalization, using the synthetic control developed here.
The calculations also suggest that \ce{Th(SO4)2} is a promising candidate for a clock with conversion electron readout.

\begin{acknowledgement}
This work was supported by NSF awards PHYS-2013011, PHY-2513134, and ARO award W911NF-11-1-0369.
This work used Bridges-2 at Pittsburgh Supercomputing Center through allocation PHY230110 from the ACCESS program, which is supported by NSF grants \#2138259, \#2138286, \#2138307, \#2137603, and \#2138296.
This work made use of the facilities of the Platform for the Accelerated Realization, Analysis, and Discovery of Interface Materials (PARADIM), which is supported by the National Science Foundation under Cooperative agreement no. DMR-2039380.

\end{acknowledgement}

\begin{suppinfo}

Details of experimental and computational methods and additional electronic structure figures are available in the supporting information. Crystallographic data have been deposited at the CCDC under deposition number 2582646. The optimized structure used in electronic structure calculations is included in VASP POSCAR format.

\end{suppinfo}

\bibliography{HM_all_refs,misc_arxiv_refs,apd-Th229}

@article{RN597,
	title        = {The crystal chemistry of four thorium sulfates},
	author       = {Albrecht, AJ and Sigmon, GE and Moore-Shay, L and Wei, R and Dawes, C and Szymanowski, J and Burns, PC},
	year         = 2011,
	journal      = {Journal of Solid State Chemistry},
	volume       = 184,
	pages        = {1591--1597},
	doi          = {10.1016/j.jssc.2011.04.024},
	type         = {Journal Article}
}

@article{RN344,
	title        = {Electronegativity values from thermochemical data},
	author       = {Allred, A. L.},
	year         = 1961,
	journal      = {Journal of Inorganic \& Nuclear Chemistry},
	volume       = 17,
	number       = {3-4},
	pages        = {215--221},
	doi          = {10.1016/0022-1902(61)80142-5},
	issn         = {0022-1902},
	url          = {<Go to ISI>://WOS:A1961WN28400004},
	type         = {Journal Article}
}

@article{RN607,
	title        = {Concept of spinning magnetic field at magic-angle condition for line narrowing in {M\"ossbauer} spectroscopy},
	author       = {Anisimov, P and Rostovtsev, Y and Kocharovskaya, O},
	year         = 2007,
	journal      = {Physical Review B},
	volume       = 76,
	doi          = {10.1103/PhysRevB.76.094422},
	type         = {Journal Article}
}

@article{RN14,
	title        = {Projector augmented wave method},
	author       = {Blochl, P. E.},
	year         = 1994,
	journal      = {Physical Review B},
	volume       = 50,
	number       = 24,
	pages        = {17953--17979},
	doi          = {10.1103/PhysRevB.50.17953},
	issn         = {1098-0121},
	type         = {Journal Article}
}

@article{RN595,
	title        = {Band-edge levels in semiconductors and insulators: Hybrid density functional theory versus many-body perturbation theory},
	author       = {Chen, W and Pasquarello, A},
	year         = 2012,
	journal      = {Physical Review B},
	volume       = 86,
    pages        = 035134,
	doi          = {10.1103/PhysRevB.86.035134},
	type         = {Journal Article}
}

@article{RN611,
	title        = {\ce{^{229}}Thorium-doped calcium fluoride for nuclear laser spectroscopy},
	author       = {Dessovic, P and Mohn, P and Jackson, RA and Winkler, G and Schreitl, M and Kazakov, G and Schumm, T},
	year         = 2014,
	journal      = {Journal of Physics - Condensed Matter},
	volume       = 26,
	doi          = {10.1088/0953-8984/26/10/105402},
	type         = {Journal Article}
}

@article{RN543,
	title        = {Investigation of Thorium Salts As Candidate Materials for Direct Observation of the \ce{^{229}Th} Nuclear Transition},
	author       = {Ellis, JK and Wen, XD and Martin, RL},
	year         = 2014,
	journal      = {Inorganic Chemistry},
	volume       = 53,
	pages        = {6769--6774},
	doi          = {10.1021/ic500570u},
	type         = {Journal Article}
}

@article{RN575,
	title        = {Laser Excitation of the \ce{^{229}Th} Nuclear Isomeric Transition in a Solid-State Host},
	author       = {Elwell, R and Schneider, C and Jeet, J and Terhune, JES and Morgan, HWT and Alexandrova, AN and Tan, HBT and Derevianko, A and Hudson, ER},
	year         = 2024,
	journal      = {Physical Review Letters},
	volume       = 133,
    pages        = 013201,
	doi          = {10.1103/PhysRevLett.133.013201},
	type         = {Journal Article}
}

@article{RN552,
	title        = {Feasibility and potential of a thorium-doped barium-lithium-fluoride single crystal as a candidate for solid-state nuclear optical clock material},
	author       = {Gong, QR and Tao, SL and Li, SM and Deng, GL and Zhao, CC and Hang, Y},
	year         = 2024,
	journal      = {Physical Review A},
	volume       = 109,
    pages        = 033109,
	doi          = {10.1103/PhysRevA.109.033109},
	type         = {Journal Article}
}

@article{RN577,
	title        = {DVM-X$_\alpha$ calculations on the ionization potentials of \ce{MX_{k+1}} complex anions and the electron affinities of \ce{MX_{k+1}} super halogens},
	author       = {Gutsev, GL and Boldyrev, AI},
	year         = 1981,
	journal      = {Chemical physics},
	volume       = 56,
	pages        = {277--283},
	type         = {Journal Article}
}

@article{RN12,
	title        = {Efficient iterative schemes for ab initio total-energy calculations using a plane-wave basis set},
	author       = {Kresse, G. and Furthmuller, J.},
	year         = 1996,
	journal      = {Physical Review B},
	volume       = 54,
	number       = 16,
	pages        = {11169--11186},
	doi          = {10.1103/PhysRevB.54.11169},
	issn         = {1098-0121},
	url          = {<Go to ISI>://WOS:A1996VT67500040},
	type         = {Journal Article}
}

@article{RN626,
	title        = {Cubic scaling \textit{GW}: Towards fast quasiparticle calculations},
	author       = {Liu, PT and Kaltak, M and Klimes, J and Kresse, G},
	year         = 2016,
	journal      = {Physical Review B},
	volume       = 94,
	number       = 16,
	doi          = {10.1103/PhysRevB.94.165109},
	type         = {Journal Article}
}

@article{RN574,
	title        = {Atomic clocks for geodesy},
	author       = {Mehlstäubler, TE and Grosche, G and Lisdat, C and Schmidt, PO and Denker, H},
	year         = 2018,
	journal      = {Reports on Progress in Physics},
	volume       = 81,
    pages        = 064401,
	doi          = {10.1088/1361-6633/aab409},
	type         = {Journal Article}
}

@article{RN573,
	title        = {High Performance Clocks and Gravity Field Determination},
	author       = {Müller, J and Dirkx, D and Kopeikin, SM and Lion, G and Panet, I and Petit, G and Visser, PNAM},
	year         = 2018,
	journal      = {Space Science Reviews},
	volume       = 214,
    pages       = 5,
	doi          = {10.1007/s11214-017-0431-z},
	type         = {Journal Article}
}

@article{RN531,
	title        = {Nuclear Excitation of the \ce{^{229}Th} Isomer via Defect States in Doped Crystals},
	author       = {Nickerson, B. S. and Pimon, M. and Bilous, P. V. and Gugler, J. and Beeks, K. and Sikorsky, T. and Mohn, P. and Schumm, T. and Pálffy, A.},
	year         = 2020,
	journal      = {Physical Review Letters},
	volume       = 125,
	number       = 3,
    pages        = 032501,
	doi          = {10.1103/PhysRevLett.125.032501},
	issn         = {0031-9007},
	url          = {<Go to ISI>://WOS:000547891900002},
	type         = {Journal Article}
}

@article{RN532,
	title        = {Driven electronic bridge processes via defect states in \ce{^{229}Th}-doped crystals},
	author       = {Nickerson, B. S. and Pimon, M. and Bilous, P. V. and Gugler, J. and Kazakov, G. A. and Sikorsky, T. and Beeks, K. and Grüneis, A. and Schumm, T. and Pálffy, A.},
	year         = 2021,
	journal      = {Physical Review A},
	volume       = 103,
	number       = 5,
    pages        = 053120,
	doi          = {10.1103/PhysRevA.103.053120},
	issn         = {2469-9926},
	url          = {<Go to ISI>://WOS:000655979900007},
	type         = {Journal Article}
}

@article{RN601,
	title        = {On the negative vertical detachment energies of \ce{SO4^{2-}},  \ce{SeO4^{2-}} , and  \ce{SO4^{2-}(H2O)}},
	author       = {Nielsen, SB and Ejsing, AM},
	year         = 2007,
	journal      = {Chemical Physics Letters},
	volume       = 441,
	pages        = {213--215},
	doi          = {10.1016/j.cplett.2007.05.037},
	type         = {Journal Article}
}

@article{RN632,
	title        = {Nuclear clocks for testing fundamental physics},
	author       = {Peik, E and Schumm, T and Safronova, MS and Pálffy, A and Weitenberg, J and Thirolf, PG},
	year         = 2021,
	journal      = {Quantum Science and Technology},
	volume       = 6,
	number       = 3,
	doi          = {10.1088/2058-9565/abe9c2},
	type         = {Journal Article}
}

@article{RN13,
	title        = {Generalized gradient approximation made simple},
	author       = {Perdew, J. P. and Burke, K. and Ernzerhof, M.},
	year         = 1996,
	journal      = {Physical Review Letters},
	volume       = 77,
	number       = 18,
	pages        = {3865--3868},
	doi          = {10.1103/PhysRevLett.77.3865},
	issn         = {0031-9007},
	url          = {<Go to ISI>://WOS:A1996VP22500044},
	type         = {Journal Article}
}

@article{RN459,
	title        = {DFT calculation of (229)thorium-doped magnesium fluoride for nuclear laser spectroscopy},
	author       = {Pimon, M. and Gugler, J. and Mohn, P. and Kazakov, G. A. and Mauser, N. and Schumm, T.},
	year         = 2020,
	journal      = {Journal of Physics-Condensed Matter},
	volume       = 32,
	number       = 25,
    pages        = 255503,
	doi          = {10.1088/1361-648X/ab7c90},
	issn         = {0953-8984},
	url          = {<Go to ISI>://WOS:000523457900001},
	type         = {Journal Article}
}

@article{RN460,
	title        = {Band Gap Calculations for Thorium-Doped \ce{LiCAF}},
	author       = {Pimon, M. and Mohn, P. and Schumm, T.},
	year         = 2022,
	journal      = {Advanced Theory and Simulations},
	volume       = 5,
	number       = 10,
    pages        = 2200185,
	doi          = {10.1002/adts.202200185},
	url          = {<Go to ISI>://WOS:000843650900001},
	type         = {Journal Article}
}

@article{RN578,
	title        = {Constraining the Evolution of the Fundamental Constants with a Solid-State Optical Frequency Reference Based on the \ce{^{229}Th} Nucleus},
	author       = {Rellergert, WG and DeMille, D and Greco, RR and Hehlen, MP and Torgerson, JR and Hudson, ER},
	year         = 2010,
	journal      = {Physical Review Letters},
	volume       = 104,
    pages        = 200802,
	doi          = {10.1103/PhysRevLett.104.200802},
	type         = {Journal Article}
}

@inproceedings{RN604,
	title        = {Progress towards fabrication of \ce{^{229}Th}-doped high energy band-gap crystals for use as a solid-state optical frequency reference},
	author       = {Rellergert, WG and Sullivan, ST and DeMille, D and Greco, RR and Hehlen, MP and Jackson, RA and Torgerson, JR and Hudson, ER},
	booktitle    = {11th Europhysical Conference on Defects in Insulating Materials (EURODIM 2010)},
	volume       = 15,
	doi          = {10.1088/1757-899X/15/1/012005},
	type         = {Conference Proceedings}
}

@article{RN627,
	title        = {Space-time method for ab-initio calculations of self-energies and dielectric response functions of solids},
	author       = {Rojas, HN and Godby, RW and Needs, RJ},
	year         = 1995,
	journal      = {Physical Review Letters},
	volume       = 74,
	number       = 10,
	pages        = {1827--1830},
	type         = {Journal Article}
}

@article{RN624,
	title        = {Implementation and performance of the frequency-dependent \textit{GW} method within the PAW framework},
	author       = {Shishkin, M and Kresse, G},
	year         = 2006,
	journal      = {Physical Review B},
	volume       = 74,
	number       = 3,
    pages        = 035101,
	doi          = {10.1103/PhysRevB.74.035101},
	type         = {Journal Article}
}

@article{RN625,
	title        = {Self-consistent \textit{GW} calculations for semiconductors and insulators},
	author       = {Shishkin, M and Kresse, G},
	year         = 2007,
	journal      = {Physical Review B},
	volume       = 75,
	number       = 23,
	doi          = {10.1103/PhysRevB.75.235102},
	type         = {Journal Article}
}

@article{RN572,
	title        = {Recent progress on the design and applications of superhalogens},
	author       = {Srivastava, AK},
	year         = 2023,
	journal      = {Chemical Communications},
	volume       = 59,
	pages        = {5943--5960},
	doi          = {10.1039/d3cc00428g},
	type         = {Journal Article}
}

@article{RN571,
	title        = {Superhalogens as Building Blocks of Ionic Liquids},
	author       = {Srivastava, AK and Kumar, A and Misra, N},
	year         = 2021,
	journal      = {Journal of Physical Chemistry A},
	volume       = 125,
	pages        = {2146--2153},
	doi          = {10.1021/acs.jpca.1c00599},
	type         = {Journal Article}
}

@article{RN544,
	title        = {Laser Excitation of the \ce{^{229}Th} Nucleus},
	author       = {Tiedau, J and Okhapkin, MV and Zhang, K and Thielking, J and Zitzer, G and Peik, E and Schaden, F and Pronebner, T and Morawetz, I and De Col, LT and Schneider, F and Leitner, A and Pressler, M and Kazakov, GA and Beeks, K and Sikorsky, T and Schumm, T},
	year         = 2024,
	journal      = {Physical Review Letters},
	volume       = 132,
    pages        = 182501,
	doi          = {10.1103/PhysRevLett.132.182501},
	type         = {Journal Article}
}

@article{RN489,
	title        = {Accurate Band Gaps of Semiconductors and Insulators with a Semilocal Exchange-Correlation Potential},
	author       = {Tran, F. and Blaha, P.},
	year         = 2009,
	journal      = {Physical Review Letters},
	volume       = 102,
	number       = 22,
    pages        = 226401,
	doi          = {10.1103/PhysRevLett.102.226401},
	issn         = {0031-9007},
	type         = {Journal Article}
}

@article{RN628,
	title        = {The GW-Method for Quantum Chemistry Applications: Theory and Implementation},
	author       = {van Setten, MJ and Weigend, F and Evers, F},
	year         = 2013,
	journal      = {Journal of Chemical Theory and Computation},
	volume       = 9,
	number       = 1,
	pages        = {232--246},
	doi          = {10.1021/ct300648t},
	type         = {Journal Article}
}

@article{RN588,
	title        = {VASPKIT: A user-friendly interface facilitating high-throughput computing and analysis using VASP code},
	author       = {Wang, V and Xu, N and Liu, JC and Tang, G and Geng, WT},
	year         = 2021,
	journal      = {Computer physics communications},
	volume       = 267,
    pages        = 108033,
	doi          = {10.1016/j.cpc.2021.108033},
	type         = {Journal Article}
}

@article{RN602,
	title        = {Frequency ratio of the \ce{^{229m}Th} nuclear isomeric transition and the \ce{^{87}Sr} atomic clock},
	author       = {Zhang, Chuankun and Ooi, Tian and Higgins, Jacob S. and Doyle, Jack F. and von der Wense, Lars and Beeks, Kjeld and Leitner, Adrian and Kazakov, Georgy A. and Li, Peng and Thirolf, Peter G. and Schumm, Thorsten and Ye, Jun},
	year         = 2024,
	journal      = {Nature},
	volume       = 633,
	number       = 8028,
	pages        = {63--70},
	doi          = {10.1038/s41586-024-07839-6},
	issn         = {1476-4687},
	url          = {https://doi.org/10.1038/s41586-024-07839-6},
	type         = {Journal Article}
}

@article{RN629,
	title        = {\ce{^229ThF4} thin films for solid-state nuclear clocks},
	author       = {Zhang, C. K. and von der Wense, L. and Doyle, J. F. and Higgins, J. S. and Ooi, T. and Friebel, H. U. and Ye, J. and Elwell, R. and Terhune, J. E. S. and Morgan, H. W. T. and Alexandrova, A. N. and Tan, H. B. T. and Derevianko, A. and Hudson, E. R.},
	year         = 2024,
	journal      = {Nature},
	volume       = 636,
	number       = 8043,
    pages        = {603--608},
	doi          = {10.1038/s41586-024-08256-5},
	type         = {Journal Article}
}

@article{Betke2011,
	title        = {Oleum and Sulfuric Acid as Reaction Media: The Actinide Examples \ce{UO2(S2O7)}‐lt (low temperature),  \ce{UO2(S2O7)}‐ht (high temperature),  \ce{UO2(HSO4)2},  \ce{An(SO4)2} ({An = Th,  U}),  \ce{Th4(HSO4)2(SO4)7} and \ce{Th(HSO4)2(SO4)}},
	author       = {Betke,  Ulf and Wickleder,  Mathias S.},
	year         = 2011,
	month        = {dec},
	journal      = {European Journal of Inorganic Chemistry},
	publisher    = {Wiley},
	volume       = 2012,
	number       = 2,
	pages        = {306–317},
	doi          = {10.1002/ejic.201100975},
	issn         = {1099-0682},
	url          = {http://dx.doi.org/10.1002/ejic.201100975}
}

@article{Morgan2025ICrate,
  title = {Theory of Internal Conversion of the \ce{^229Th} Nuclear Isomer in Solid-State Hosts},
  volume = {134},
  pages = 253801,
  ISSN = {1079-7114},
  url = {http://dx.doi.org/10.1103/9s8f-hv1f},
  DOI = {10.1103/9s8f-hv1f},
  number = {25},
  journal = {Physical Review Letters},
  publisher = {American Physical Society (APS)},
  author = {Morgan,  H. W. T. and Tran Tan,  H. B. and Elwell,  R. and Alexandrova,  A. N. and Hudson,  Eric R. and Derevianko,  Andrei},
  year = {2025},
  month = jun 
}

@article{Guo2016ThS2,
	title        = {Structural and electronic phase transitions of \ce{ThS2} from first-principles calculations},
	author       = {Guo,  Yongliang and Wang,  Changying and Qiu,  Wujie and Ke,  Xuezhi and Huai,  Ping and Cheng,  Cheng and Zhu,  Zhiyuan and Chen,  Changfeng},
	year         = 2016,
	month        = {oct},
	journal      = {Physical Review B},
	publisher    = {American Physical Society (APS)},
	volume       = 94,
	number       = 13,
	doi          = {10.1103/physrevb.94.134104},
	issn         = {2469-9969},
	url          = {http://dx.doi.org/10.1103/PhysRevB.94.134104}
}

@article{Griffiths2000ThO2,
	title        = {Electron irradiation of single crystal thorium dioxide and electron transfer reactions},
	author       = {Griffiths,  Trevor R and Dixon,  James},
	year         = 2000,
	month        = {apr},
	journal      = {Inorganica Chimica Acta},
	publisher    = {Elsevier BV},
	volume       = {300–302},
	pages        = {305–313},
	doi          = {10.1016/s0020-1693(99)00597-6},
	issn         = {0020-1693},
	url          = {http://dx.doi.org/10.1016/S0020-1693(99)00597-6}
}

@article{Pereira2017ThO2,
	title        = {Optical properties of \ce{ThO2}–based nanoparticles},
	author       = {Pereira,  F.J. and Castro,  M.A. and V\'azquez,  M.D. and Deb\'an,  L. and Aller,  A.J.},
	year         = 2017,
	month        = {apr},
	journal      = {Journal of Luminescence},
	publisher    = {Elsevier BV},
	volume       = 184,
	pages        = {169–178},
	doi          = {10.1016/j.jlumin.2016.12.025},
	issn         = {0022-2313},
	url          = {http://dx.doi.org/10.1016/j.jlumin.2016.12.025}
}

@misc{Morgan2025ThSO42,
	title        = {A spinless crystal for a high-performance solid-state $^{229}$Th nuclear clock},
	author       = {Morgan,  Harry W. T. and Terhune,  James E. S. and Elwell,  Ricky and Tan,  Hoang Bao Tran and Perera,  Udeshika C. and Derevianko,  Andrei and Hudson,  Eric R. and Alexandrova,  Anastassia N.},
	year         = 2025,
	publisher    = {arXiv},
	doi          = {10.48550/ARXIV.2503.11374},
	url          = {https://arxiv.org/abs/2503.11374},
	copyright    = {Creative Commons Attribution Non Commercial Share Alike 4.0 International},
	eprint       = {2408.12309},
	archiveprefix = {arXiv}
}

@article{Pineda2025,
	title        = {Radiative decay of the \ce{^{229m}Th} nuclear clock isomer in different host materials},
	author       = {Pineda,  S. V. and Chhetri,  P. and Bara,  S. and Elskens,  Y. and Casci,  S. and Alexandrova,  A. N. and Au,  M. and Athanasakis-Kaklamanakis,  M. and Bartokos,  M. and Beeks,  K. and Bernerd,  C. and Claessens,  A. and Chrysalidis,  K. and Cocolios,  T. E. and Correia,  J. G. and De Witte,  H. and Elwell,  R. and Ferrer,  R. and Heinke,  R. and Hudson,  E. R. and Ivandikov,  F. and Kudryavtsev,  Yu. and K\"{o}ster,  U. and Kraemer,  S. and Laatiaoui,  M. and Lica,  R. and Merckling,  C. and Morawetz,  I. and Morgan,  H. W. T. and Moritz,  D. and Pereira,  L. M. C. and Raeder,  S. and Rothe,  S. and Schaden,  F. and Scharl,  K. and Schumm,  T. and Stegemann,  S. and Terhune,  J. and Thirolf,  P. G. and Tunhuma,  S. M. and Van Den Bergh,  P. and Van Duppen,  P. and Vantomme,  A. and Wahl,  U. and Yue,  Z.},
	year         = 2025,
	month        = {jan},
	journal      = {Physical Review Research},
	publisher    = {American Physical Society (APS)},
	volume       = 7,
	number       = 1,
    pages        = 013052,
	doi          = {10.1103/physrevresearch.7.013052},
	issn         = {2643-1564},
	url          = {http://dx.doi.org/10.1103/PhysRevResearch.7.013052}
}

@article{Morgan2025polyatomic,
  title = {Design of new thorium nuclear clock materials based on polyatomic ions},
  ISSN = {1477-9234},
  url = {http://dx.doi.org/10.1039/D5DT00736D},
  volume = 54,
  pages = {10574-10580},
  DOI = {10.1039/d5dt00736d},
  journal = {Dalton Transactions},
  publisher = {Royal Society of Chemistry (RSC)},
  author = {Morgan,  Harry W. T. and Tan,  Hoang Bao Tran and Derevianko,  Andrei and Elwell,  Ricky and Terhune,  James E. S. and Hudson,  Eric R. and Alexandrova,  Anastassia N.},
  year = {2025}
}

@article{Knope2011,
  title = {Synthesis and Characterization of Thorium(IV) Sulfates},
  volume = {50},
  ISSN = {1520-510X},
  url = {http://dx.doi.org/10.1021/ic201175u},
  DOI = {10.1021/ic201175u},
  number = {17},
  journal = {Inorganic Chemistry},
  publisher = {American Chemical Society (ACS)},
  author = {Knope,  Karah E. and Wilson,  Richard E. and Skanthakumar,  S. and Soderholm,  L.},
  year = {2011},
  month = aug,
  pages = {8621–8629}
}

@article{Elwell2025ThO2,
  title = {Laser-based conversion electron M\"{o}ssbauer spectroscopy of \ce{^229ThO2}},
  volume = {648},
  ISSN = {1476-4687},
  url = {http://dx.doi.org/10.1038/s41586-025-09776-4},
  DOI = {10.1038/s41586-025-09776-4},
  number = {8093},
  journal = {Nature},
  publisher = {Springer Science and Business Media LLC},
  author = {Elwell,  Ricky and Terhune,  James E. S. and Schneider,  Christian and Morgan,  Harry W. T. and Tan,  Hoang Bao Tran and Perera,  Udeshika C. and Rehn,  Daniel A. and Alfonso,  Marisa C. and von der Wense,  Lars and Seiferle,  Benedict and Scharl,  Kevin and Thirolf,  Peter G. and Derevianko,  Andrei and Hudson,  Eric R.},
  year = {2025},
  month = dec,
  pages = {300–305}
}

@misc{Hiraki2025LaserMossbauer,
  doi = {10.48550/ARXIV.2509.00041},
  url = {https://arxiv.org/abs/2509.00041},
  author = {Hiraki,  Takahiro and Masuda,  Takahiko and Takatori,  Sayuri and Schaden,  Fabian and Bartokos,  Michael and Beeks,  Kjeld and Fukunaga,  Yuta and Gr\"{u}neis,  Andreas and Guan,  Ming and Kazakov,  Georgy and LaGrange,  Thomas and Leitner,  Adrian and Morawetz,  Ira and Ogake,  Ryoichiro and Okai,  Koichi and Pimon,  Martin and Pressler,  Martin and Riebner,  Thomas and Sasao,  Noboru and Schneider,  Felix and Schumm,  Thorsten and Shimizu,  Kotaro and de Col,  Luca Toscani and Sikorsky,  Tomas and Yoshimi,  Akihiro and Yoshimura,  Koji},
  title = {Laser M\"{o}ssbauer spectroscopy of \ce{^229Th}},
  publisher = {arXiv},
  year = {2025},
  copyright = {Creative Commons Attribution 4.0 International}
}

@article{Gatti2013,
  title = {Exciton dispersion from first principles},
  volume = {88},
  ISSN = {1550-235X},
  url = {http://dx.doi.org/10.1103/PhysRevB.88.155113},
  DOI = {10.1103/physrevb.88.155113},
  number = {15},
  journal = {Physical Review B},
  publisher = {American Physical Society (APS)},
  author = {Gatti,  Matteo and Sottile,  Francesco},
  year = {2013},
  month = Oct 
}

@article{Cheng2022,
  title = {Direct-determination of high-concentration sulfate by serial differential spectrophotometry with multiple optical pathlengths},
  volume = {811},
  ISSN = {0048-9697},
  url = {http://dx.doi.org/10.1016/j.scitotenv.2021.152121},
  DOI = {10.1016/j.scitotenv.2021.152121},
  journal = {Science of The Total Environment},
  publisher = {Elsevier BV},
  author = {Cheng,  Wen and Zhang,  Xuefei and Duan,  Ning and Jiang,  Linhua and Xu,  Yanli and Chen,  Ying and Liu,  Yong and Fan,  Peng},
  year = {2022},
  month = Mar,
  pages = {152121}
}

@article{Cheng2023,
  title = {The characteristics of ultraviolet absorption and electronic excitation of sulfate at high concentrations},
  volume = {293},
  ISSN = {1386-1425},
  url = {http://dx.doi.org/10.1016/j.saa.2023.122455},
  DOI = {10.1016/j.saa.2023.122455},
  journal = {Spectrochimica Acta Part A: Molecular and Biomolecular Spectroscopy},
  publisher = {Elsevier BV},
  author = {Cheng,  Wen and Duan,  Ning and Jiang,  Linhua and Xu,  Yanli and Zhu,  Guangbin and Zhang,  Xuefei and Liu,  Yong and Chen,  Ying and Zhang,  Rong and Xu,  Fuyuan},
  year = {2023},
  month = May,
  pages = {122455}
}

@article{Ai2021,
  title = {A monitoring method for sulfate based on ultraviolet absorption spectroscopy dedicated to SO3 monitoring in coal-fired power plants},
  volume = {780},
  ISSN = {0009-2614},
  url = {http://dx.doi.org/10.1016/j.cplett.2021.138935},
  DOI = {10.1016/j.cplett.2021.138935},
  journal = {Chemical Physics Letters},
  publisher = {Elsevier BV},
  author = {Ai,  Xinyu and Li,  Leipeng and Zhou,  Xue and Zhang,  Zhiguo},
  year = {2021},
  month = Oct,
  pages = {138935}
}

@article{Ai2022,
  title = {Quantitative measurement model for sulfate in the temperature range of 298.15–343.15 K based on vacuum ultraviolet absorption spectroscopy},
  volume = {200},
  ISSN = {0042-207X},
  url = {http://dx.doi.org/10.1016/j.vacuum.2022.111035},
  DOI = {10.1016/j.vacuum.2022.111035},
  journal = {Vacuum},
  publisher = {Elsevier BV},
  author = {Ai,  Xinyu and Zhang,  Zhiguo},
  year = {2022},
  month = June,
  pages = {111035}
}

@article{Pyykk2014,
  title = {Additive Covalent Radii for Single-,  Double-,  and Triple-Bonded Molecules and Tetrahedrally Bonded Crystals: A Summary},
  volume = {119},
  ISSN = {1520-5215},
  url = {http://dx.doi.org/10.1021/jp5065819},
  DOI = {10.1021/jp5065819},
  number = {11},
  journal = {The Journal of Physical Chemistry A},
  publisher = {American Chemical Society (ACS)},
  author = {Pyykk\"{o},  Pekka},
  year = {2014},
  month = Sept,
  pages = {2326–2337}
}

@book{Fox2010optical,
  title     = "Optical properties of solids",
  author    = "Fox, Mark",
  publisher = "Oxford University Press",
  series    = "Oxford Master Series in Physics",
  edition   =  2,
  month     =  mar,
  year      =  2010,
  address   = "London, England"
}

@misc{Derevianko2026Colloquium,
  title = {Colloquium: {{Nuclear}} Clocks},
  shorttitle = {Colloquium},
  author = {Derevianko, Andrei and Elwell, R. and Hudson, Eric R.},
  year = 2026,
  month = jun,
  number = {arXiv:2606.11048},
  eprint = {2606.11048},
  primaryclass = {physics.atom-ph},
  publisher = {arXiv},
  doi = {10.48550/arXiv.2606.11048},
   url = {https://arxiv.org/abs/2606.11048},
  urldate = {2026-06-10},
  archiveprefix = {arXiv}
}

@misc{toscanidecol2026clock,
  doi = {10.48550/ARXIV.2606.04997},
  url = {https://arxiv.org/abs/2606.04997},
  author = {Toscani De Col,  L. and Riebner,  T. and Morawetz,  I. and Schneider,  F. and Sempelmann,  N. and Schlachet-Lépinay,  J. and Schaden,  F. and Bartokos,  M. and Kazakov,  G. A. and Beeks,  K. and Gerstenecker,  B. and Pimon,  M. and Lahs,  S. and Hellerschmied,  A. and Lercher,  T. and Premper,  J. and Niessner,  A. and Matus,  M. and Denker,  H. and Cizek,  M. and Cip,  O. and Lal,  V. and Zitzer,  G. and Petrov,  V. and Tiedau,  J. and Okhapkin,  M. V. and Peik,  E. and Schumm,  T.},
  title = {A thorium-229 optical nuclear clock with feedback loop},
  publisher = {arXiv},
  year = {2026},
  copyright = {Creative Commons Attribution 4.0 International}
}

@misc{huang2026clock,
  doi = {10.48550/ARXIV.2606.08870},
  url = {https://arxiv.org/abs/2606.08870},
  author = {Huang,  Beichen and Yan,  Gaowei and Xiao,  Qi and Bu,  Wenhao and Zhang,  Zhen and Zhao,  Chengchun and Yan,  Chao and Chen,  Zhi-Ang and Zhang,  Peixiong and Penyazkov,  Gleb and Zhan,  Zhenhai and Yan,  Lingfeng and Wang,  Yuefei and Li,  Lin and Li,  Shanming and Qian,  Xiaobo and Liu,  Xuegang and He,  Qiange and Sun,  Taoxiang and Tian,  Haochen and Lu,  Binkun and Ma,  Ningyuan and Li,  Juxian and Wu,  Yanzhang and Gong,  Qiaorui and Li,  Yuxiang and Shi,  Haoyu and Li,  Xiangliang and Ma,  Longsheng and Zhu,  Shining and Mo,  Yuxiang and Lin,  Jun and You,  Li and Lin,  Yige and Zhang,  Xibo and Hang,  Yin and Su,  Liangbi and Ding,  Shiqian},
  title = {A nuclear clock based on \ce{^229Th}},
  publisher = {arXiv},
  year = {2026},
  copyright = {arXiv.org perpetual,  non-exclusive license}
}

@article{Ladenburg1933,
  author  = {Ladenburg, R. and Van Voorhis, C. C.},
  title   = {The Continuous Absorption of Oxygen Between 1750 and 1300~{\AA} and Its Bearing upon the Dispersion},
  journal = {Physical Review},
  year    = {1933},
  volume  = {43},
  number  = {5},
  pages   = {315--321},
  doi     = {10.1103/PhysRev.43.315}
}

@article{Serralheiro2015,
  author  = {Serralheiro, C. and Duflot, D. and Ferreira da Silva, F. and Hoffmann, S. V. and Jones, N. C. and Mason, N. J. and Mendes, B. and Lim{\~a}o-Vieira, P.},
  title   = {Toluene Valence and {R}ydberg Excitations as Studied by ab Initio Calculations and Vacuum Ultraviolet ({VUV}) Synchrotron Radiation},
  journal = {The Journal of Physical Chemistry A},
  year    = {2015},
  volume  = {119},
  number  = {34},
  pages   = {9059--9069},
  doi     = {10.1021/acs.jpca.5b05080}
}

@article{Nickerson2021,
  title = {Driven Electronic Bridge Processes via Defect States in {{229Th-doped}} Crystals},
  author = {Nickerson, Brenden S and Pimon, Martin and Bilous, Pavlo V and Gugler, Johannes and Kazakov, Georgy A and Sikorsky, Tomas and Beeks, Kjeld and Gr{\"u}neis, Andreas and Schumm, Thorsten and P{\'a}lffy, Adriana},
  year = 2021,
  journal = {Phys. Rev. A},
  volume = {103},
  number = {5},
  pages = {053120},
  issn = {2469-9926},
  doi = {10.1103/physreva.103.053120}
}

@misc{Perera2026AbInitio,
  type = {Preprint},
  title = {Ab Initio Calculations of {$^{229}$}{{Th}} Band-to-Band Internal Conversion Rate in {$^{229}$}{{ThO}}{$_{2}$}},
  author = {Perera, Udeshika C. and Tran Tan, H. B. and Morgan, H. W. T. and Hudson, Eric and Rehn, Daniel A. and Derevianko, Andrei},
  year = 2026,
  month = jul,
  number = {arXiv:2607.08941},
  eprint = {2607.08941},
  primaryclass = {nucl-th},
  publisher = {arXiv},
  doi = {10.48550/arXiv.2607.08941},
  urldate = {2026-07-13},
  archiveprefix = {arXiv}
}

@misc{BaMgF4,
  doi = {10.48550/ARXIV.2410.23364},
    eprint={2410.23364},
      archivePrefix={arXiv},
  url = {https://arxiv.org/abs/2410.23364},
  author = {Morgan,  H. W. T. and Elwell,  R. and Terhune,  J. E. S. and Tan,  H. B. Tran and Perera,  U. C. and Derevianko,  A. and Alexandrova,  A. N. and Hudson,  E. R.},
  title = {\ce{^{229}Th}-doped nonlinear optical crystals for compact solid-state clocks},
  publisher = {arXiv},
  year = {2024},
  copyright = {Creative Commons Attribution Non Commercial Share Alike 4.0 International}
}

@misc{IC_rate_theory,
  doi = {10.48550/ARXIV.2411.15641},
  url = {https://arxiv.org/abs/2411.15641},
   eprint={2411.15641},
      archivePrefix={arXiv},
  author = {Morgan,  H. W. T. and Tan,  H. B. Tran and Elwell,  R. and Alexandrova,  A. N. and Hudson,  Eric R. and Derevianko,  Andrei},
  title = {Theory of internal conversion of the thorium-229 nuclear isomer in solid-state hosts},
  publisher = {arXiv},
  year = {2024},
  copyright = {Creative Commons Attribution 4.0 International}
}

@misc{our_photoquenching,
      title={Photo-Induced Quenching of the \ce{^229Th} Isomer in a Solid-State Host}, 
      author={J. E. S. Terhune and R. Elwell and H. B. Tran Tan and U. C. Perera and H. W. T. Morgan and A. N. Alexandrova and Andrei Derevianko and Eric R. Hudson},
      year={2024},
      eprint={2412.08998},
      archivePrefix={arXiv},
      primaryClass={physics.atom-ph},
      url={https://arxiv.org/abs/2412.08998}, 
}

@misc{PTB_photoquenching,
      title={Laser-Induced Quenching of the {Th-229} Nuclear Clock Isomer in Calcium Fluoride}, 
      author={F. Schaden and T. Riebner and I. Morawetz and L. Toscani De Col and G. A. Kazakov and K. Beeks and T. Sikorsky and T. Schumm and K. Zhang and V. Lal and G. Zitzer and J. Tiedau and M. V. Okhapkin and E. Peik},
      year={2024},
      eprint={2412.12339},
      archivePrefix={arXiv},
      primaryClass={physics.atom-ph},
      url={https://arxiv.org/abs/2412.12339}, 
}

@misc{dynamical_decoupling,
      title={Dynamical decoupling in interacting systems: applications to signal-enhanced hyperpolarized readout}, 
      author={A. Ajoy and R. Nirodi and A. Sarkar and P. Reshetikhin and E. Druga and A. Akkiraju and M. McAllister and G. Maineri and S. Le and A. Lin and A. M. Souza and C. A. Meriles and B. Gilbert and D. Suter and J. A. Reimer and A. Pines},
      year={2020},
      eprint={2008.08323},
      archivePrefix={arXiv},
      primaryClass={quant-ph},
      url={https://arxiv.org/abs/2008.08323}, 
}

\end{document}



\title{Investigation into thorium sulfate as a spinless crystal for a high-performance solid-state \ce{^229Th} nuclear clock \\ Supplementary information}

\author{Harry W. T. Morgan}
\affiliation{Department of Chemistry, University of Manchester, Oxford Road, Manchester M13 9PL, UK}
\affiliation{Department of Chemistry and Biochemistry, University of California, Los Angeles, Los Angeles, CA 90095, USA}

\author{James E. S. Terhune}
\author{Albert Bao}
\author{Ricky Elwell}
\affiliation{Department of Physics and Astronomy, University of California, Los Angeles, CA 90095, USA}

\author{Tyler Kerr}
\affiliation{Department of Chemistry and Biochemistry, University of California, Los Angeles, Los Angeles, CA 90095, USA}

\author{Alanna Iwanicki}
\author{Tyrel McQueen}
\affiliation{Department of Chemistry, Johns Hopkins University, Baltimore, MD 21218, USA}

\author{Hoang Bao Tran Tan}
\affiliation{Department of Physics, University of Nevada, Reno, Nevada 89557, USA}
\affiliation{Los Alamos National Laboratory, P.O. Box 1663, Los Alamos, New Mexico 87545, USA} 

\author{Udeshika C. Perera}
\author{Andrei Derevianko}
\affiliation{Department of Physics, University of Nevada, Reno, Nevada 89557, USA}

\author{Eric R. Hudson}
\affiliation{Department of Physics and Astronomy, University of California, Los Angeles, CA 90095, USA}

\author{Anastassia N. Alexandrova}
\affiliation{Department of Chemistry and Biochemistry, University of California, Los Angeles, Los Angeles, CA 90095, USA}

\date{\today}

\maketitle


\section{Sample Preparation}
\begin{figure}[]
    \centering
    \includegraphics[width=\linewidth]{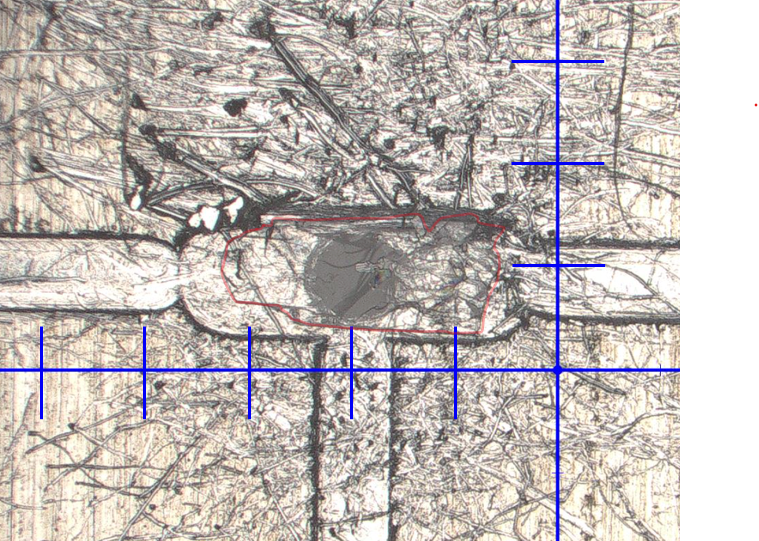}
    \caption{TM8 crystal in mount under cover slip. Crystal sits in shallow slot milled into the steel mount. Each blue tick represents 250 \textmu m. Crystal outlined in red. Scratch marks surrounding the slot are from previous loading attempts.}
    \label{fig:TM8_mount}
\end{figure}

The \ce{Th(SO4)2} crystals were stored in a vial of sulfolane. To prepare a crystal for spectroscopy, all crystals were laid out on a Petri dish filled with sulfolane and observed with a microscope. The largest and most optically transparent crystals were selected for use and transferred to the mount with a pipette. One of those crystals was then manoeuvred into a pill-shaped slot in the mount using fine tools (Fig. \ref{fig:TM8_mount}). The sample was secured beneath a \ce{CaF2} cover slip held in place with vacuum-compatible PTFE tape. T-shaped vent holes drilled into the sides of the slot  allowed trapped gas to escape during chamber pump-down. The sample mounting was carried out as quickly as possible in an effort to minimize the amount of water vapor exposure on the samples.

\section{External absorption calculations}
It is conceivable that \ce{O2} trapped between the \ce{Th(SO4)2} crystal and \ce{CaF2} cover slip could account for some portion of the observed transmission. Using the O$_2$ absorption cross section from \cite{Ladenburg1933} and an estimated maximum absorption path length of 4~\textmu m (based on the mount, crystal, and cover slip geometry), we calculate a 1/$e$ absorption onset near 145~nm. Since this is well short of the 180~nm onset observed in our transmission spectrum, O$_2$ absorption cannot account for the measured feature.

The first batch of crystals was originally stored and washed in toluene. After a period of roughly five days, the quality of these crystals had degraded significantly. It appeared that toluene was being either absorbed or adsorbed by the crystals. With toluene absorption cross section data from \cite{Serralheiro2015}, we calculated that toluene layers on the order of 10 nm can significantly affect our VUV absorption. To prevent this, samples were stored in sulfolane rather than toluene and washed with isopropyl alcohol three times before analysis.

\section{X-ray Diffraction}
Single-crystal X-ray diffraction (XRD) was performed on select \ce{Th(SO4)2} samples from the same batch as the crystal whose transmission spectrum is plotted in Fig. 8 of the main text. Measurements at 100~K showed an orthorhombic unit cell with lengths $a = 9.5099(7)$~\AA, $b = 9.0909(6)$~\AA, and $c = 13.624(1)$~\AA. 3D reconstructions from the XRD data also confirmed the structure as \ce{Th(SO4)2}, verifying our transmission measurement as that of \ce{Th(SO4)2}.

\section{Electronic structure calculations}

Calculations were performed with VASP~\cite{RN12,RN14}, version 6.4.2 unless otherwise specified.
The crystal structure of \ce{Th(SO4)2} was optimized using PBE~\cite{RN13} functional, with a $k$-point spacing of 0.03 \AA{}$^{-1}$ (4-3-2 Monkhorst-Pack mesh) and a 500 eV plane wave cutoff.
The optimized lattice parameters ($a = 9.65511$~\AA, $b = 9.26933$~\AA, $c = 13.88311$~\AA) agree well with those measured for our sample by X-ray diffraction, validating the computational model.
The optimized structure was used for subsequent electronic structure calculations.
$k$-point grids were generated with VASPKIT~\cite{RN588}.

The band gap of \ce{Th(SO4)2} was computed with the $G_0W_0R$ method as implemented in VASP~\cite{RN624,RN625,RN626,RN627}.
The method is similar to the non-self-consistent $G_0W_0$ in which orbitals are computed with DFT and then energy corrections are computed in a single iteration of $GW$.
$G_0W_0R$ differs from $G_0W_0$ in that the latter scales as $N^4$ while the former scales as $N^3$ but has higher memory requirements.
$G_0W_0R$ is therefore suitable for larger systems.

When calculating the band gap of \ce{Th(SO4)2} with $G_0W_0R$ we tested parameters in the method to ensure that the results were not sensitive to them.
These parameters include the number of frequency and time grid points used in the calculation (NOMEGA), the DFT method used to generate the orbitals, and the pseudopotentials for sulfur and oxygen.
A specialized $GW$ pseudopotential for thorium is not available.
The results of these tests are shown in Table~\ref{G0W0R parameter results table}.
The important conclusions are that the band gap is converged when NOMEGA = 16, the gap is not sensitive to the choice of DFT orbitals, and the $GW$ pseudopotentials do not make a significant difference to the gap for this system.
In the main text we report the band gap of 8.18 eV that was computed with NOMEGA = 16, $GW$ pseudopotentials for O and S, and PBE orbitals.
For all calculations in the main part of the study the plane wave cut-off was 400 eV and the $k$-point spacing was 0.05 \AA{}$^{-1}$ (2-2-1 $\Gamma$-centered mesh).

\begin{table}[h]
\begin{ruledtabular}
\begin{tabular}{c|c|c}
        & NOMEGA        & Band gap (eV)        \\
        \hline
PBE orbitals, normal PPs & &  \\
        & 10            & 8.00                 \\
        & 12            & 8.07                 \\
        & 16            & 8.07                 \\
        \hline
PBE orbitals, GW PPs & &      \\
        & 12            & 8.17                 \\
        & 16            & 8.18                 \\
        \hline
LDA orbitals, GW PPs & &     \\
        & 12            & 8.11                
\end{tabular}
\end{ruledtabular}
\caption{Band gap of \ce{Th(SO4)2} computed with $G_0W_0R$ with different parameters. ``PP'' stands for ``pseudopotential'' and denotes whether the standard or $GW$-specialized pseudopotentials were used for oxygen and sulfur.}
\label{G0W0R parameter results table}
\end{table}

Increasing the plane wave cut-off to 450 eV increased the band gap to 8.24 eV, which we consider to be an insignificant difference compared to 8.18 eV.
A calculation with a cut-off of 500 eV was attempted but the calculation became too expensive.

Increasing the $k$-mesh to 2-2-2 gave a band gap of 8.17 eV, from which we conclude that a 2-2-1 $k$-mesh is sufficient.
A calculation with a 3-3-2 mesh was attempted but the calculation was too expensive.

All calculations agree that the valence band maximum is at $\Gamma$.
In some calculations, the conduction band minimum is also at $\Gamma$, giving a direct band gap, and the lowest energy orbital at $X$ is degenerate.
In other calculations the orbital at $X$ is marginally lower ($\sim 0.01$ eV), giving a formally indirect band gap.

BSE calculations used the Tamm-Dancoff approximation and read 2000 bands from the $G_0W_0R$ step.
All other parameters were the same as in the $G_0W_0R$ step.
Excitations were computed with 128 occupied bands, 192 unoccupied bands, and a maximum excitation energy of 14 eV.

\begin{figure}[h]
    \centering
    \includegraphics[width=0.8\linewidth]{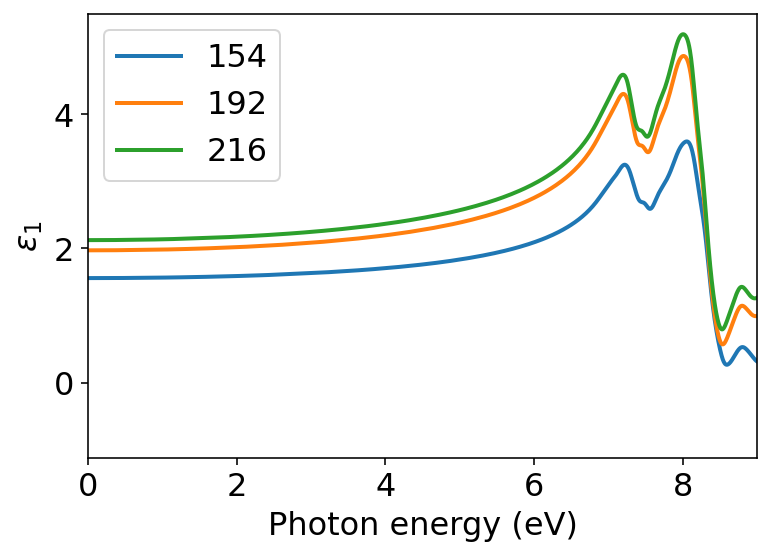}
    \caption{Convergence of $\epsilon_1$ with respect to the number of unoccupied states in the BSE calculation. Each curve is labelled by its NBANDSV value.}
    \label{fig:e1_convergence}
\end{figure}

\begin{figure}[h]
    \centering
    \includegraphics[width=0.8\linewidth]{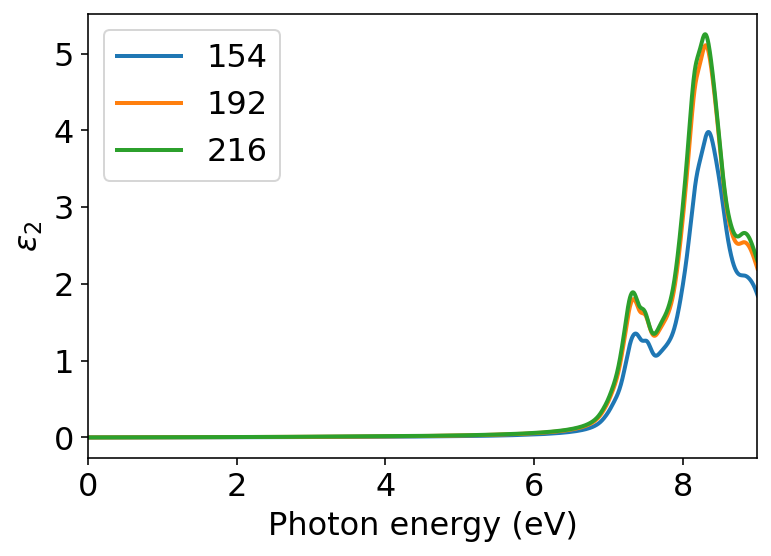}
    \caption{Convergence of $\epsilon_2$ with respect to the number of unoccupied states in the BSE calculation. Each curve is labelled by its NBANDSV value.}
    \label{fig:e2_convergence}
\end{figure}

One possible explanation for the difference between the experimental sample thickness and that used for the transmission spectrum calculations is that $\epsilon_1$ (the real part of the dielectric function) is not converged with respect to the number of unoccupied states (NBANDSV) or maximum excitation energy (OMEGAMAX) in the BSE step.
That explanation is indicated by the fact that the transmission below 180 nm is severely underestimated when $d=10\; \mu m$ - in this range $\epsilon_2$ (the imaginary part of the dielectric function) is zero, so any predicted absorption must come from $\epsilon_1$.
Convergence was tested by varying the number of unoccupied states and maximum excitation energy as indicated in Table \ref{LREAL convergence parameter table}.
The shape of $\epsilon_1$ is constant across the calculations, as shown in Fig. \ref{fig:e1_convergence}, and the quantitative differences are small suggesting that non-convergence is not the origin of the thickness mismatch.
For completeness, $\epsilon_2$ plots from the same calculations are shown in Figure \ref{fig:e2_convergence}.

\begin{table}[h]
\begin{ruledtabular}
\begin{tabular}{c|c|c}
 NBANDSV       & OMEGAMAX (eV)        & \# transitions        \\
        \hline
 154       & 12            & 27732                 \\
 192       & 14            & 58930                \\
 216       & 16            & 81805                 
\end{tabular}
\end{ruledtabular}
\caption{Parameters used to test convergence of $\epsilon_1$ with respect to the number of empty states in the BSE calculation}
\label{LREAL convergence parameter table}
\end{table}

\section{Exciton plots}

BSE calculations for exciton plotting were done with VASP version 6.5.1 due to errors in this functionality in earlier versions.

The exciton wavefunction has the form:
\begin{align}
    \psi(r_e,r_h) &= \sum_{vc} A_{vc} \phi_v(r_e)\phi_c(r_h)
\end{align}
where $r_e$ and $r_h$ are respectively the positions of the electron and the hole in the excition, $\phi_v$ and $\phi_c$ are respectively the valence and conduction band orbitals, and $A_{vc}$ is the excitonic Hamiltonian matrix element that couples pairs of orbitals (one occupied, one unoccupied) to form the exciton.\cite{Gatti2013}

Crucially, the excitonic wavefunction $\psi(r_e,r_h)$ is a function of the coordinates of the electron and the hole.
To plot the real-space distribution of the electron, we have to fix the coordinate of the hole, or \textit{vice versa}.
Fixing a coordinate requires chemical understanding of the system.
In our case, we can be confident that the hole is located on an O atom because the O lone pairs are the highest-energy electrons in the system.
We therefore fix the hole within 1 pm of an O nucleus (for technical reasons they cannot have identical coordinates).

\begin{figure}[htb]
    \centering
    \includegraphics[width=\linewidth]{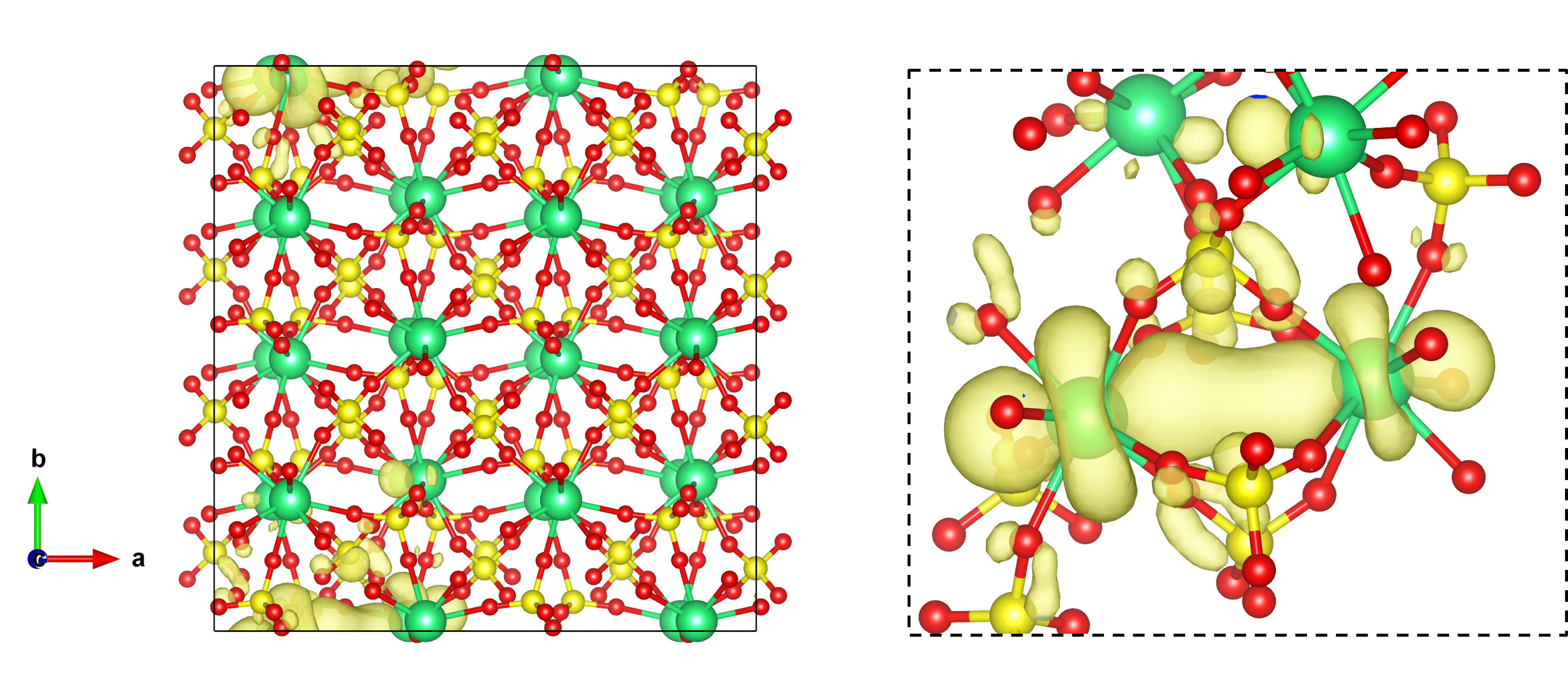}
    \caption{Excitonic electron density for the 7.28 eV transition (oscillator strength = 1.43) with the hole on O.}
    \label{fig:CHG.40 plot}
\end{figure}

\begin{figure}[htb]
    \centering
    \includegraphics[width=\linewidth]{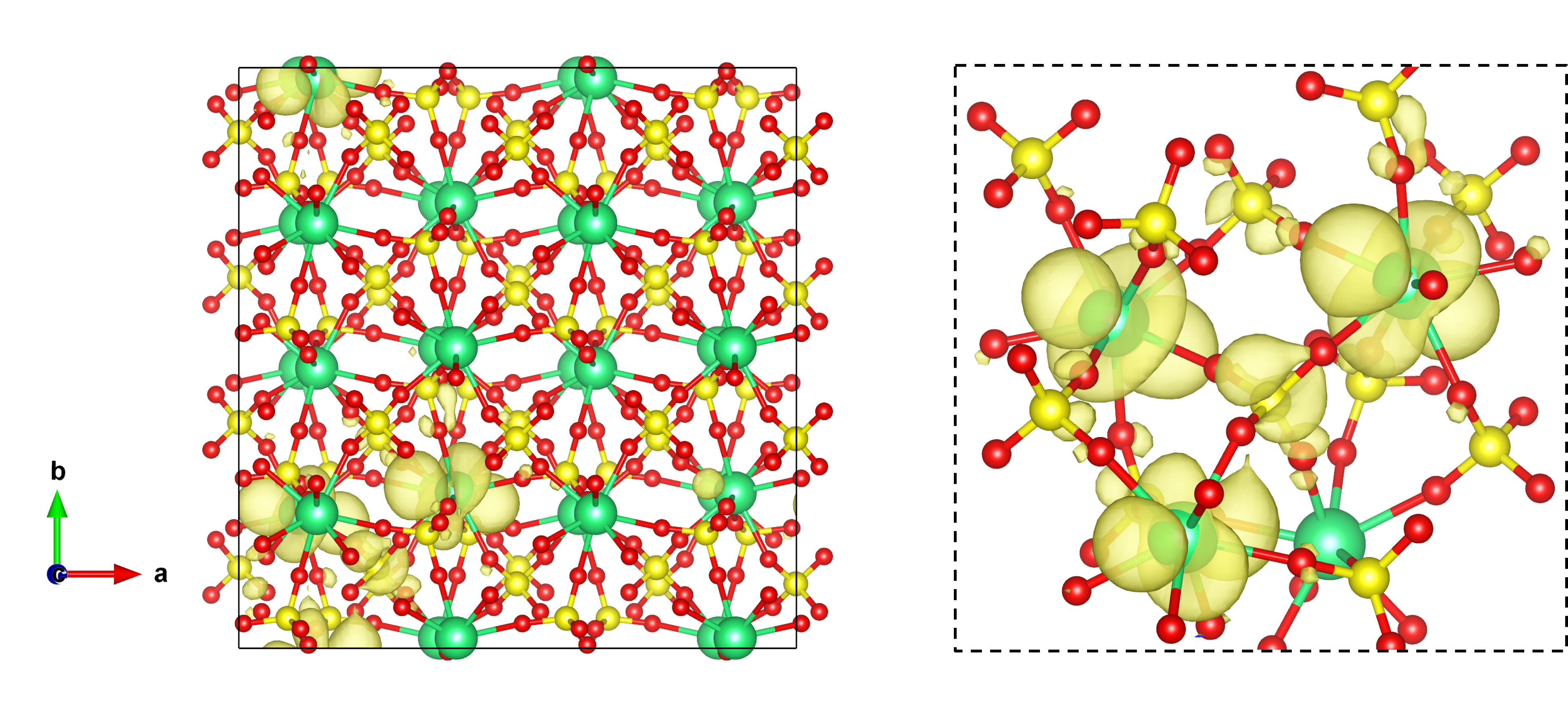}
    \caption{Excitonic electron density for the 7.29 eV transition (oscillator strength = 1.19) with the hole on O.}
    \label{fig:CHG.44 plot}
\end{figure}

\begin{figure}[htb]
    \centering
    \includegraphics[width=\linewidth]{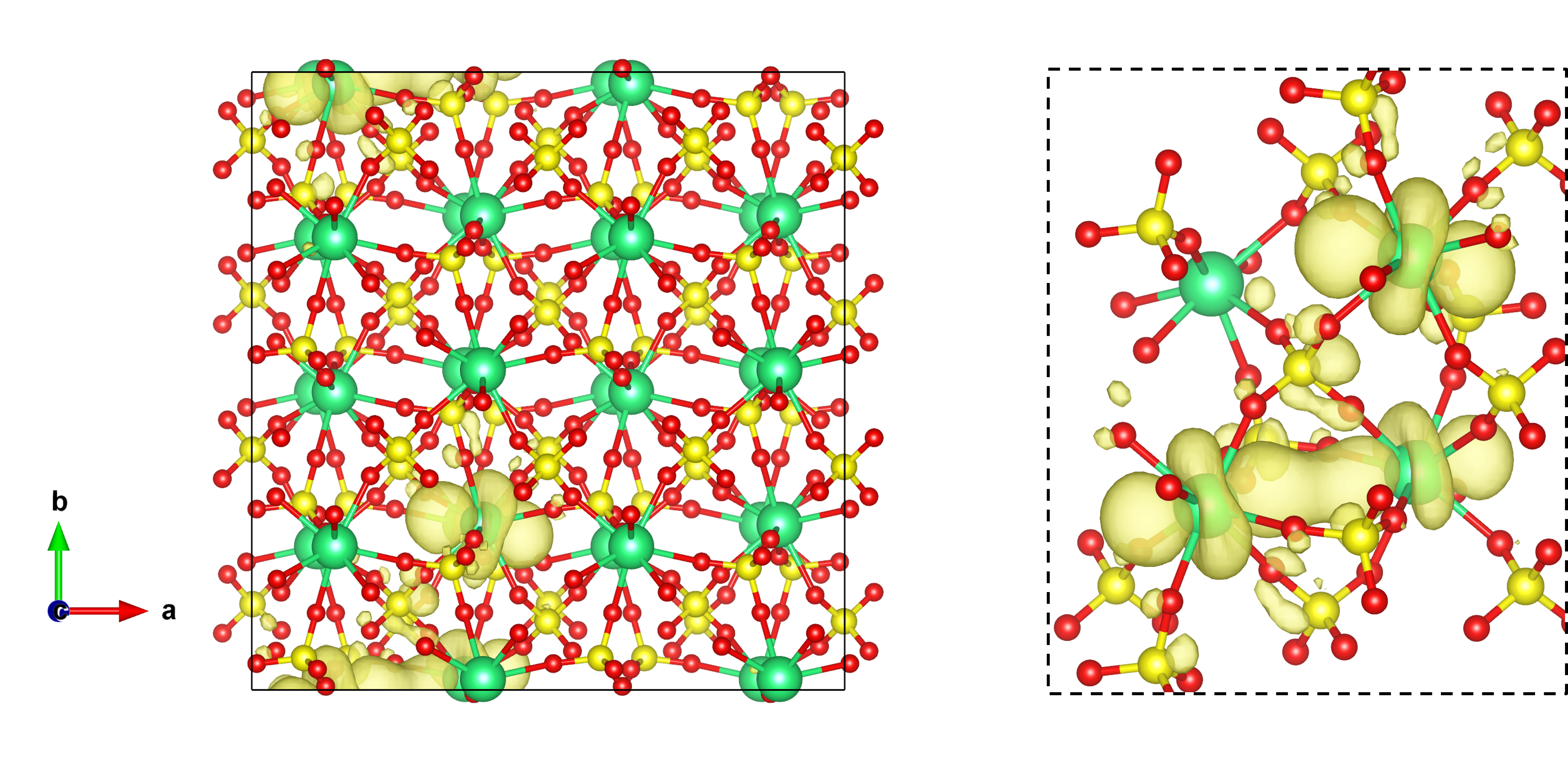}
    \caption{Excitonic electron density for the 7.34 eV transition (oscillator strength = 1.00) with the hole on O.}
    \label{fig:CHG.51 plot}
\end{figure}

As stated in the main text, transitions with the highest oscillator strengths are at 7.28 eV, 7.29 eV, 7.34 eV, and 7.36 eV.
The 7.36 eV excitonic electron density is shown in the main text; the others are in Figures \ref{fig:CHG.40 plot}-\ref{fig:CHG.51 plot}.

All have predominantly Th $d$ orbital character, so we can characterize the transitions as LMCT (ligand-to-metal charge transfer).
Three of the four plots (all except 7.29 eV) show clear $d-d$ orbital overlap between neighbouring thorium atoms, while the 7.29 eV plot shows three Th $d$ orbitals mixed together with a central \ce{SO4^{2-}} $\pi^*$ orbital - the phases can be inferred from normal $d$ orbitals.

\begin{figure}[htb]
    \centering
    \includegraphics[width=\linewidth]{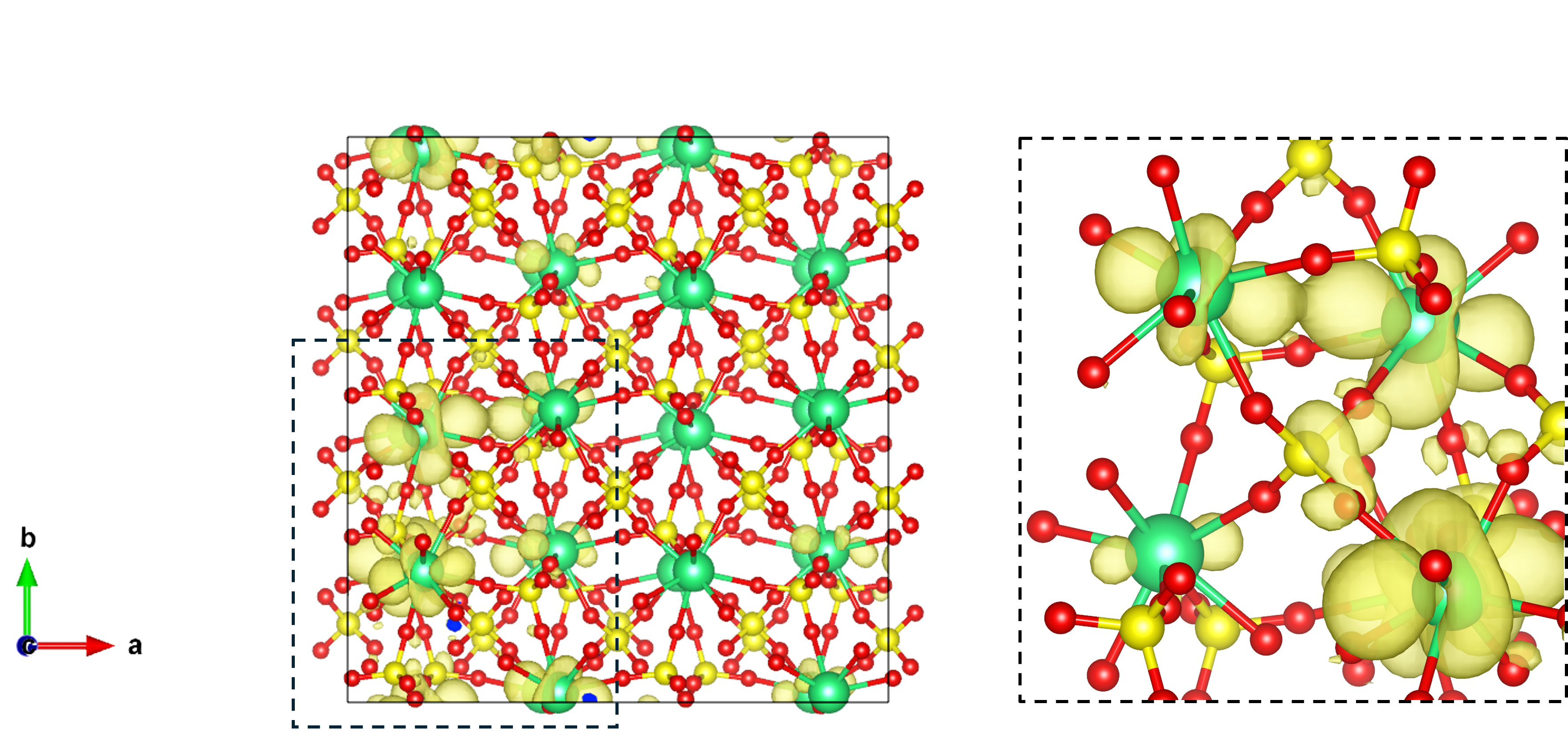}
    \caption{Excitonic electron density for the 7.28 eV transition (oscillator strength = 1.43) with the hole on S.}
    \label{fig:S hole CHG.40 plot}
\end{figure}

\begin{figure}[htb]
    \centering
    \includegraphics[width=\linewidth]{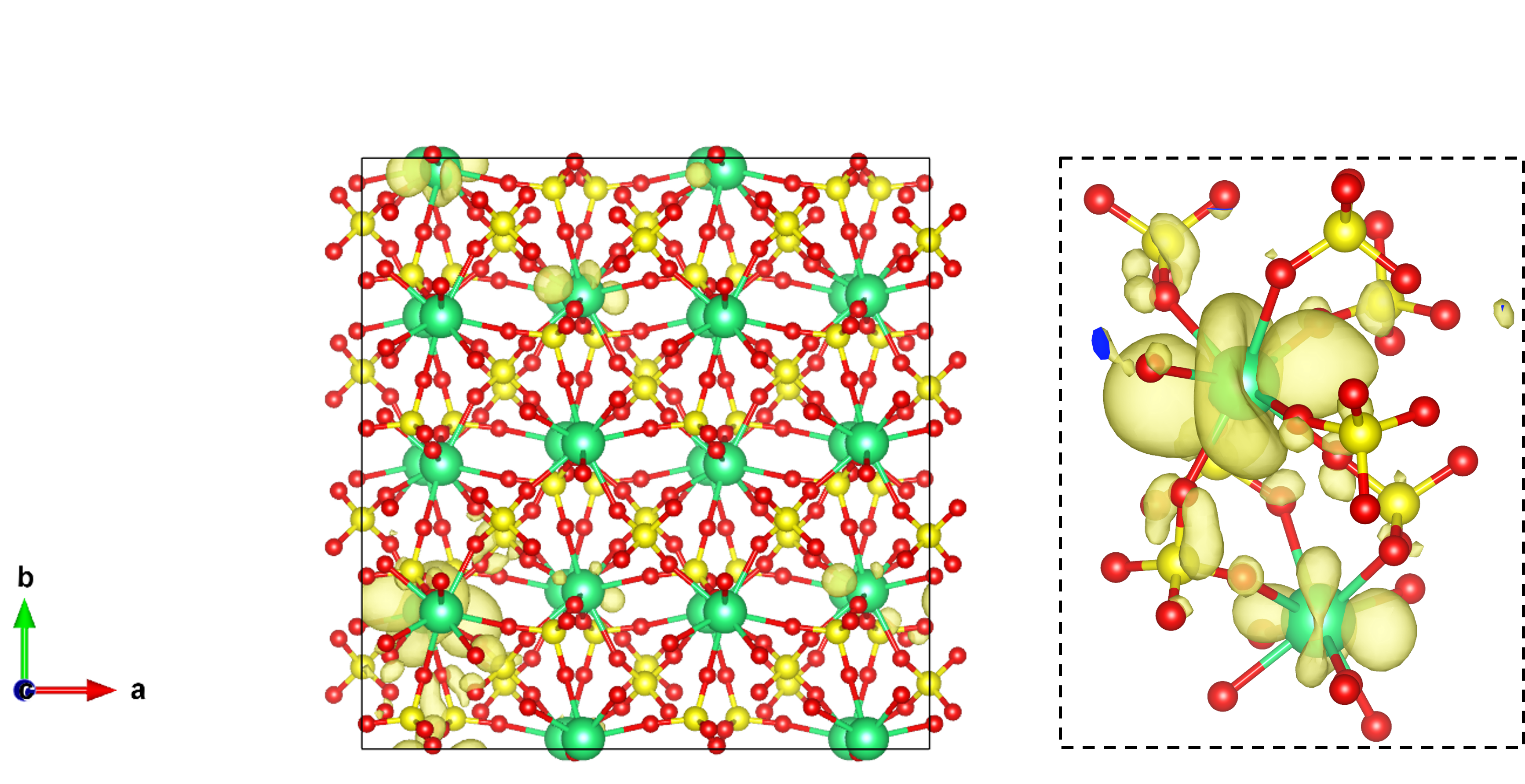}
    \caption{Excitonic electron density for the 7.29 eV transition (oscillator strength = 1.19) with the hole on S.}
    \label{fig:S hole CHG.44 plot}
\end{figure}

\begin{figure}[htb]
    \centering
    \includegraphics[width=\linewidth]{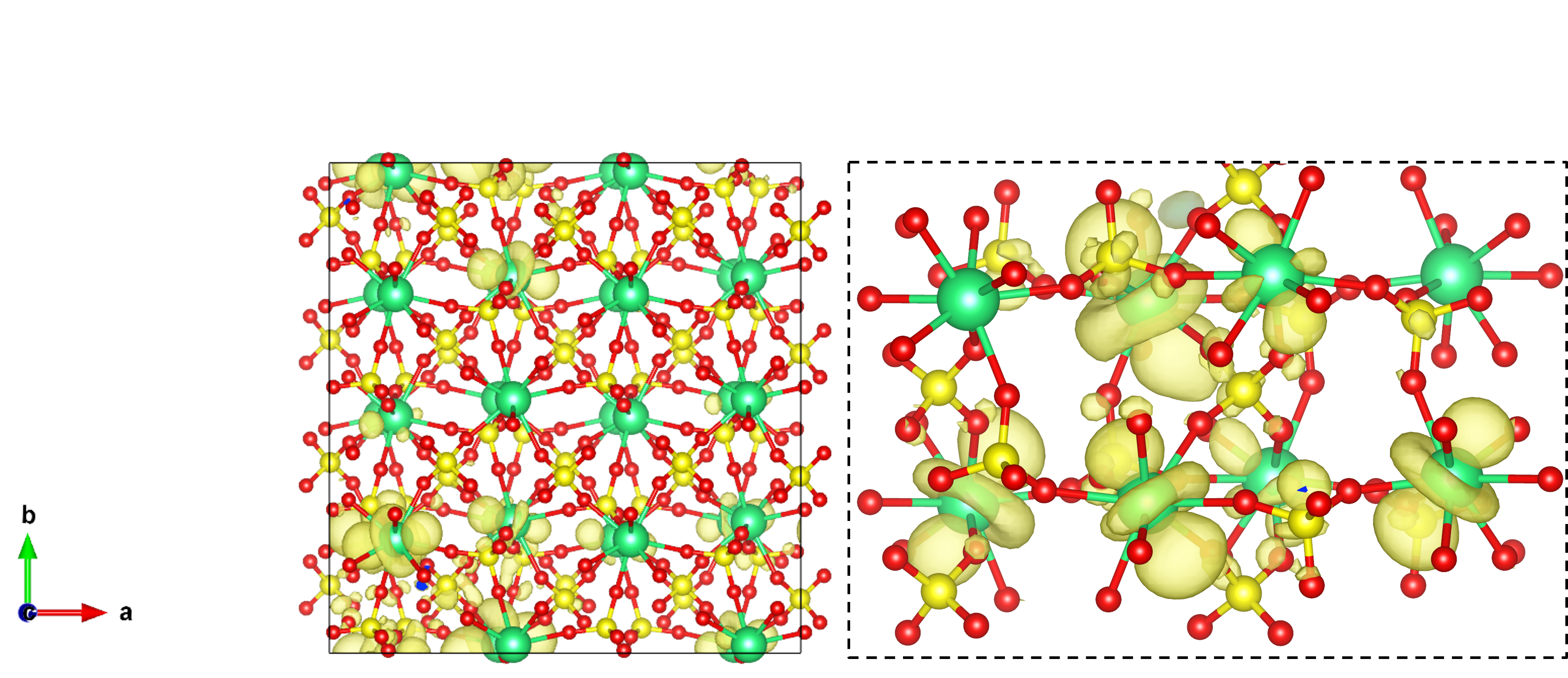}
    \caption{Excitonic electron density for the 7.34 eV transition (oscillator strength = 1.00) with the hole on S.}
    \label{fig:S hole CHG.51 plot}
\end{figure}

\begin{figure}[htb]
    \centering
    \includegraphics[width=\linewidth]{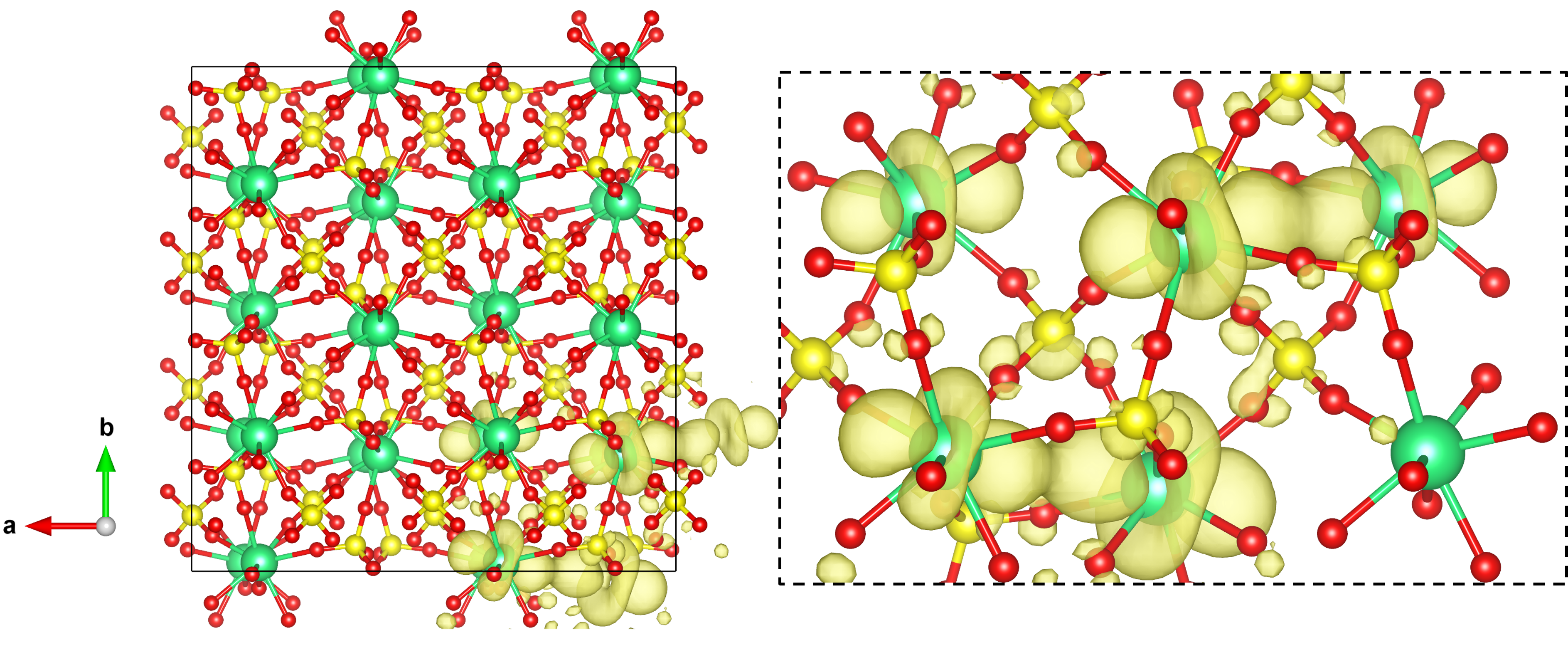}
    \caption{Excitonic electron density for the 7.36 eV transition (oscillator strength = 1.47) with the hole on S.}
    \label{fig:S hole CHG.56 plot}
\end{figure}

To check the robustness of the results we also plot the excitonic electron densities for the same transitions but with the hole fixed close to an S nucleus.
The plots are shown in Figures \ref{fig:S hole CHG.40 plot}-\ref{fig:S hole CHG.56 plot}.
The electron densities are still predominantly Th $d$ in nature, featuring some $d-d$ overlaps, but the densities are spread over more Th atoms.


\bibliography{HM_all_refs,misc_arxiv_refs,ab_references}